\documentclass[twocolumn]{aastex63}
\usepackage{ulem}
\usepackage{color}
\usepackage{lipsum}
\usepackage{amsmath}
\usepackage{mathtools}
\usepackage{graphics}
\usepackage{comment}
\usepackage{graphicx}
\usepackage[justification=centering, font=footnotesize]{caption}
\usepackage{enumerate}
\usepackage{mdframed}
\usepackage{layouts}
\usepackage{sidecap}
\usepackage{xcolor}

\newcommand{\tauColl}{\tau_{\mbox{coll}}}
\newcommand{\etal}{et al}
\submitjournal{AJ}

\shortauthors{Savitch et al.}
\shorttitle{Triggering A Climate Change Dominated "Anthropocene": Is It Common Among Exocivilizations?}

\begin{document}

\title{Triggering A Climate Change Dominated "Anthropocene": Is It Common Among Exocivilizations?}

\author[0000-0002-2919-1109]{Ethan Savitch}
\author[0000-0002-4948-7820]{Adam Frank}
\author[0000-0003-3265-7210]{Jonathan Carroll-Nellenback}
\affiliation{Department of Physics and Astronomy, University of Rochester, Rochester, New York, 14620}

\author[0000-0003-4346-2611]{Jacob Haqq-Misra}
\affiliation{Blue Marble Space Institute of Science, 600 1st Avenue, 1st Floor, Seattle, Washington 98104}

\author[0000-0002-3798-0730]{Axel Kleidon}
\affiliation{Max-Planck-Institut fuer Biogeochemie, Jena, Germany}

\author[0000-0002-1920-309X]{Marina Alberti}
\affiliation{Department of Urban Design and Planning, University of Washington, Seattle, WA 98195, USA}

\begin{abstract}
    We seek to model the coupled evolution of a planet and a civilization through the era when energy harvesting by the civilization drives the planet into new and adverse climate states. In this way we ask if triggering "anthropocenes" of the kind humanity is experiencing now might be a generic feature of planet-civilization evolution.  In this study we focus on the effects of energy harvesting via combustion and vary the planet's initial atmospheric chemistry and orbital radius. In our model, energy harvesting increases the civilization's population growth rate while also, eventually, leading to a degradation of the planetary climate state (relative to the civilization's habitability.) We also assume the existence of a Complex Life Habitable Zone in which very high levels of $CO_2$ are detrimental to multi-cellular animal life such as those creating technological civilizations.  Our models show that the civilization's growth is truncated by planetary feedback (a "climate dominated anthropocene") for a significant region of the initial parameter space.
\end{abstract}


\section{Introduction}
It has now become clear that human activity has altered the state of the coupled Earth systems (atmosphere, hydrosphere, cryosphere, lithosphere, biosphere). There are multiple measures of human impact on these systems including the transport of key compounds and materials \citep{Lenton2015}; the colonization of surface area \citep{Hooke2013}; human appropriation of the terrestrial productivity \citep{VitousekEtal1986} and energy \citep{Kleidon2012}.  Global Warming driven by $CO_2$ emissions represents the most dramatic example of the impact of civilization on the planet \citep{Solomon2007}. 

Taken as a whole these changes in the state/behavior of Earth's coupled planetary systems have been described as a new planetary/geologic epoch termed the {\it Anthropocene} \citep{Crutzen2002}.  In fact, recent studies have shown that 2020 marks the moment when human-made 'anthropogenic' mass has exceeded all of Earth's living biomass \citep{Elahacham2020}.  The specifics of the long-term impact of the Anthropocene on human civilization is difficult to predict. These impacts are, however, accepted to have negative consequences with assessments ranging from a difficult adaptation to full-scale collapse. Also unknown are the requirements needed to successfully manage our entry into the Anthropocene and then create a long-term sustainable version of civilization.  One can even ask if such long-term sustainable versions of civilization are even possible.  It is possible that the Anthropocene may represent a ``tipping point" in the coupled dynamical system representing both planet and civilization such that once the point is crossed in state space, subsequent evolution proves detrimental to the civilization. \citep{Lenton2008,Kuehn2011}. We note that in \cite{Frank2017} final sustainable planetary states were explored. 

In \cite{Frank2018} the Earth's entry into the Anthropocene was examined from an astrobiological perspective. That study asked if the situation currently encountered on Earth was unique.  In particular, given its global scale, might the transition represented by the Anthropocene be a generic feature of any planet evolving a species that intensively harvests resources for the development of a technological civilization \citep{Haqq-Misra2009,Frank2017,MullanHaqq-Misra2019}?  This question has direct consequences for both the study of astrobiology and the sustainability of human civilization. 

Relevant to astrobiology, it is now apparent that most stars harbor families of planets \citep{Seager2013}. Indeed, many of those planets will be in the star's habitable zones \citep{Howard2013}. Tremendous effort has gone into the study of biosignatures, i.e. imprints a biosphere leaves on detectable light from the planet. Recently it has been recognized that imprints from technology created by an intelligent civilization might be just as, or more easily detectable \citep{LingamLoeb2019} than "traditional" biosignatures.  If Anthropocenes are a common consequence of a civilization developing on a given inhabited world, then this co-evolutionary period between planet and civilization may effect the nature, and even existence, of technosignatures. In addition, if anthropocenes prove fatal for some civilizations then they can be considered as one form of a "Great Filter" and are therefore relevant to discussions of the Fermi Paradox \citep{Carroll-Nellenbackea2019}

The possibility that anthropocenes are common is equally of interest to the pressing concerns about our own immediate future. We are, essentially, without a playbook in dealing with the planetary transition we now face (though however \cite{Frank2017}). Any understanding of generic features in the co-evolution of planetary systems and civilizations could be of use in charting out the possible futures for our own efforts to navigate our own version of the Anthropocene. Even purely modeling/theoretical perspectives on how techno-spheres co-evolve with the other geospheres (the biosphere in particular) may help us understand the range and efficacy of viable options.

The modeling framework presented in \cite{Frank2018} was meant as a first step in studying generic behaviors in the interaction between a resource-harvesting technological civilization (an {\it exo-civilization}) and the planetary environment in which it evolves. Using methods from dynamical systems theory, a suite of simple equations was introduced for modeling a population which consumes resources (for the purpose of running a technological civilization) and the feedback those resources drive on the state of the host planet.  The feedbacks drive the planet away from the initial state that gave birth to the civilization. The simple models in \cite{Frank2018} conceptualized the problem primarily in terms of feedbacks from the resource use onto the coupled planetary systems, including "population growth advantages" gained via the harvesting of the resources. The models showed three distinct classes of exo-civilization trajectories. The first of these were smooth entries into long-term, ``sustainable" steady-states.  The second class were population booms followed by various levels of ``die-off".  Finally were rapid ``collapse" trajectories for which the population ($N$) approaches $N = 0$.  

In this work we seek to take a step up in complexity and realism compared to \cite{Frank2018}.  In particular, we represent the evolution of the planetary state via an explicit energy balance climate model (EBM) and take the global temperature to be representative of that state \citep{NorthKim2017}.  The interaction between the civilization and the planetary coupled systems is mediated by the civilization's $CO_2$ production.  This means we are explicitly considering civilizations whose energy generation comes through some form of combustion. As in the first paper, the use of this energy allows the civilization to increase its population (via increases in the birth rate of the population).  At the same time, the feedback of the energy use on the planetary state, now via $CO_2$ emissions, alters that state. Thus planetary conditions can be driven beyond what is tolerable for the functioning of the civilization.  This is reflected in an increase in the mortality (the death rate) of the population.

We explore the effect of changing two key parameters in the models: the orbital distance of the planet from its host star and the initial atmospheric chemistry of the planet in terms of $CO_2$.  In essence we are asking if we moved Earth to different orbits and/or changed its initial $CO_2$ concentration, would we still have triggered the climate change we are experiencing now.  

Most animal life on Earth can not tolerate high levels of $CO_2$ \citep{WittmannPortner2013}.  Thus a key assumption of our models is that complex life of the kind expected to build a technological civilization will emerge from a "Complex Life Habitable Zone" (CLHZ) where initial $CO_2$ concentrations are below a threshold \citep{Schwietermanea2019,Ramirez2020,Catlingea2005}.  We will discuss the consequences of this assumption in the discussion section. 

Finally we emphasize that this paper focuses on the {\it triggering} of a anthropocene which we will define as planetary-systems change created by the civilization which truncates the civilization's population growth. When the main driver of the end of population growth is climate change for brevity we call this a "climate-dominated anthropocene". This is in contrast to asking what a civilization can do to manage such an anthropocene once it occurs.  To deal with this second question requires including the civilization's response (and the timing of that response) in the model.  This is a topic which should be addressed in future work.

The plan of the paper is as follows. In section 2 we describe the model and use the Earth's recent history to tune and test it. In section 3 we provide an analysis of a linearized version of the model to extract key features of its behavior in terms of dimensionless parameters. In section 4 we present the results of the full non-linear model. We first show two sets of experiments run with either constant initial temperature or constant initial atmospheric $CO_2$ concentration ($pCO_2$). We then show the results from a sweep of two parameters: orbital distance ($a$) and initial ($pCO_2$). Finally we run a suite of models in which the civilization tolerance for global temperature changes ($\Delta T$) is varied. In section 6 we discuss the consequences of these results for emerging studies of the "astrobiology of the anthropocene" and present our conclusions and summary.

\section{The Model}

We take a dynamical systems approach to the coupled evolution of the planet and civilization. The planet is described in terms of its atmospheric state given by its average temperature $T$.  This state depends on the influx of stellar radiation and the atmospheric chemical composition which will change due to the activity of the civilization.  In our model all the civilization's energy harvesting occurs via combustion.  Thus we follow the emission of $CO_2$ by the civilization.  Changes in its partial pressure, $P(t) = pCO_2(t)$, represent the principle evolutionary driver occurring in the planet's atmospheric composition.

We use a "1-D" energy balance model (EBM) to calculate the temperature in latitudinal ($\theta$) bands.  
\begin{equation}
\frac{dT(\theta,P)}{dt}=\frac{\psi(1-A)-I+\nabla \cdot \left (\kappa \nabla  T(\theta) \right)}{C_{v}}
\label{ME1}
\end{equation}
where $A$ is the planetary albedo, $\kappa$ is the latitudinal heat transport and $C_v$ is the heat capacity at constant volume.  Our version of the model was originally developed by Darren Williams in \cite{Kastings1997}.  It was then modified by Jacob Haqq-Misra, who used it most recently in \cite{Fairen2012}.  This version of the model is publicly available on GitHub at \url{https://github.com/BlueMarbleSpace/hextor/releases/tag/1.2.2}.  In our implementation of the model we average across latitudinal bands to obtain a single globally averaged temperature.

\begin{figure}
    \centering
    \includegraphics[width=\columnwidth]{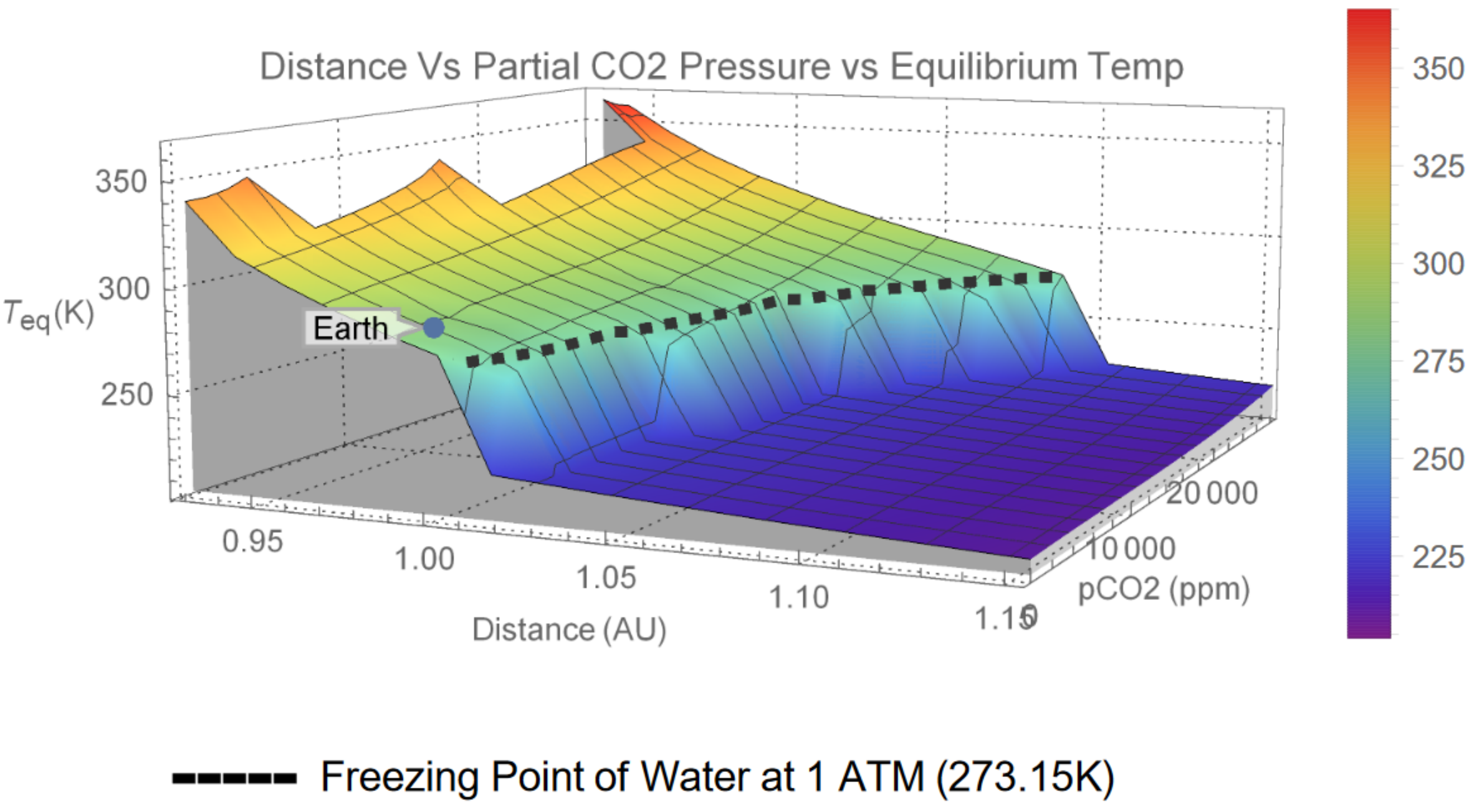}
    \caption{Surface plot of Sun's Habitable Zone, calculated with the EBM given by equation $(\ref{ME1})$.  The blue dot marks the location of Earth.}
    \label{HabZone}
\end{figure}

In Figure \ref{HabZone} we show the domains of our model in $(a,P_0)$ space where $P_0= pCO_{2,0}$ is the initial $CO_2$ composition of the planet before the civilization appears.  Note that  variations of the inner edge ($a_i$) of the habitable zone with $P_0$ are due to fits in the absorption coefficients used in the EBM.  While these could be reconciled with more detailed models, the small variation imposed on $a_{i}$ did not effect the conclusions of the study.
The dynamics of the civilization's population, $N$, is governed by the per-capita net growth rate $R$, which we let vary with time ($R=R(t)$) and define as the balance between the per-capita birth ($A$) and death ($B$) rates.  In our simulations we assume a "pre-technological" growth rate. 
\begin{equation}
R_0\equiv A_0-B_0=\frac{1}{N}\frac{dN}{dt}\bigg|_{t=t_0}
\end{equation}
As the civilization becomes more proficient at energy harvesting, it's ability to produce more offspring increases. In our model we assume the civilization's technological capacity (and hence its ability to harvest energy) tracks with the production of combustion byproducts.  In other words, growth rates depend on technology, and we take the rise in $P$ to be a measure of technological advance. Thus we define an {\it enhanced growth coefficient} to be a function of $P$ relative to the initial value $(P_0)$ the civilization found the planet in when it began its technological evolution.  For our enhanced growth coefficient we choose the form,

\begin{equation}
R_+=R_{0}\left(1+\frac{P-P_{0}}{\Delta P}\right)
\label{ME2a}
\end{equation}

Where $\Delta P$ is a normalization constant that roughly corresponds to the $pCO_2$ required to double the growth rate of the civilization.  In this way equation $(\ref{ME2a}$) captures in a simple way how increases in technology (measured by combustion products released into the atmosphere) increase the birth rate of the civilization. 

As technology produces higher $P$ and more births there will, eventually, be a corresponding feedback on the planet, dictated by equation $(\ref{ME1})$, and hence on the population.  We model this feedback via a term we denote the {\it diminished growth rate}, which we take to be temperature dependent.

\begin{equation}
R_-=R_{0}\left(\frac{T-T_{0}}{\Delta T}\right)^{2}
\label{ME2b}
\end{equation}

Where $T_{0}$ is the average planetary temperature when the civilization began ($t=t_0$), and $\Delta T$ describes the range of temperatures amenable to the civilization's health. Thus we call $\Delta T$ the civilizations {\it temperature tolerance}. This term can refer to both the biology of individuals or the functioning of the civilization as a whole.  Thus while individual members of the civilization may be able to survive at $T>T_0+\Delta T$ the civilizations functioning as a complex system may be compromised.

The final governing equation for $N$ is,

\begin{equation}
\frac{dN}{dt}=min\left[NR_+,\ R_{0}(N_{max}-N)\right] - NR_-
\label{ME2}
\end{equation}

The use of the $min$ function in equation $(\ref{ME2})$ introduces a carrying capacity ($N_{max}$) into the systems dynamics. Carrying capacity is a foundational principle in population dynamics \citep{Kot2001}.  Without it the civilization's population can grow to levels that are unrealistic based purely on food production capacities. For example, for Earth, $N > $ 100 billion appears unrealistic under even the most optimistic scenarios \citep{Cohen1995}.  In the classic logistic growth model,

\begin{equation}
\frac{dN}{dt}=NR\left(1-\frac{N}{N_{max}}\right)
\label{logeq}
\end{equation}

the carrying capacity appears in the second term which functions as the death rate.  In our model we chose to impose the carrying capacity through the $min$ function to avoid the arbitrary non-linear dependence on population which occurs in the logistic equation. We will discuss the behavior this produces in the results section. 

Finally, we model the production of $CO_2$ via the simple equation,

\begin{equation}
\frac{dP}{dt}=CN
\label{ME3}
\end{equation}

We do not include any means of reducing the $CO_2$ in the atmosphere.  While this can occur through natural means via weather and carbonate cycles, the relevant timescales are much longer than we are interested in here ($t \sim 10^6$ y).  We are also not attempting to model the possible responses of a civilization to the climate change they generate. Here we only wish to know how broad are the conditions that can lead to such change and its detrimental impacts. In terms of our equations this means there is no equilibrium for the temperature except for the trivial one of the absence of a technological civilization ($N = 0$). 

\subsection{Modeling Anthropocene Earth}
In order to provide both a test and a calibration of our model, we apply it to the recent co-evolution of Earth and its human inhabitants into the Anthropocene.  It is worth noting a few points about the initial parameters. 

The model was begun at $t=t_0 = 1820$ CE.  The global world population was taken to be $N = N_0 = 1.29\times 10^9$ with an initial $CO_2$ partial pressure of $P = P_0 = 284$ ppm, approximately equal to the values on Earth prior to the industrial revolution.  We chose the civilizations temperature tolerance of $\Delta T = 5 K$ as this is representative of the range of temperatures considered in IPCC models and acts to quantify our civilizations "fragility" \citep{Solomon2007}.  Also, our choice for the technology birth benefit ($\Delta P = 30 ppm$) was chosen as an order of magnitude approximation to the levels of $pCO_2$ humans thrive in \citep{Schwietermanea2019}.  The $CO_2$ generation coefficient $(C)$ was taken from current global conditions while the initial net growth rate $(R_0)$ was tuned to reproduce a best fit to the data.

\begin{figure}[h!]
    \centering
    \includegraphics[width=\columnwidth]{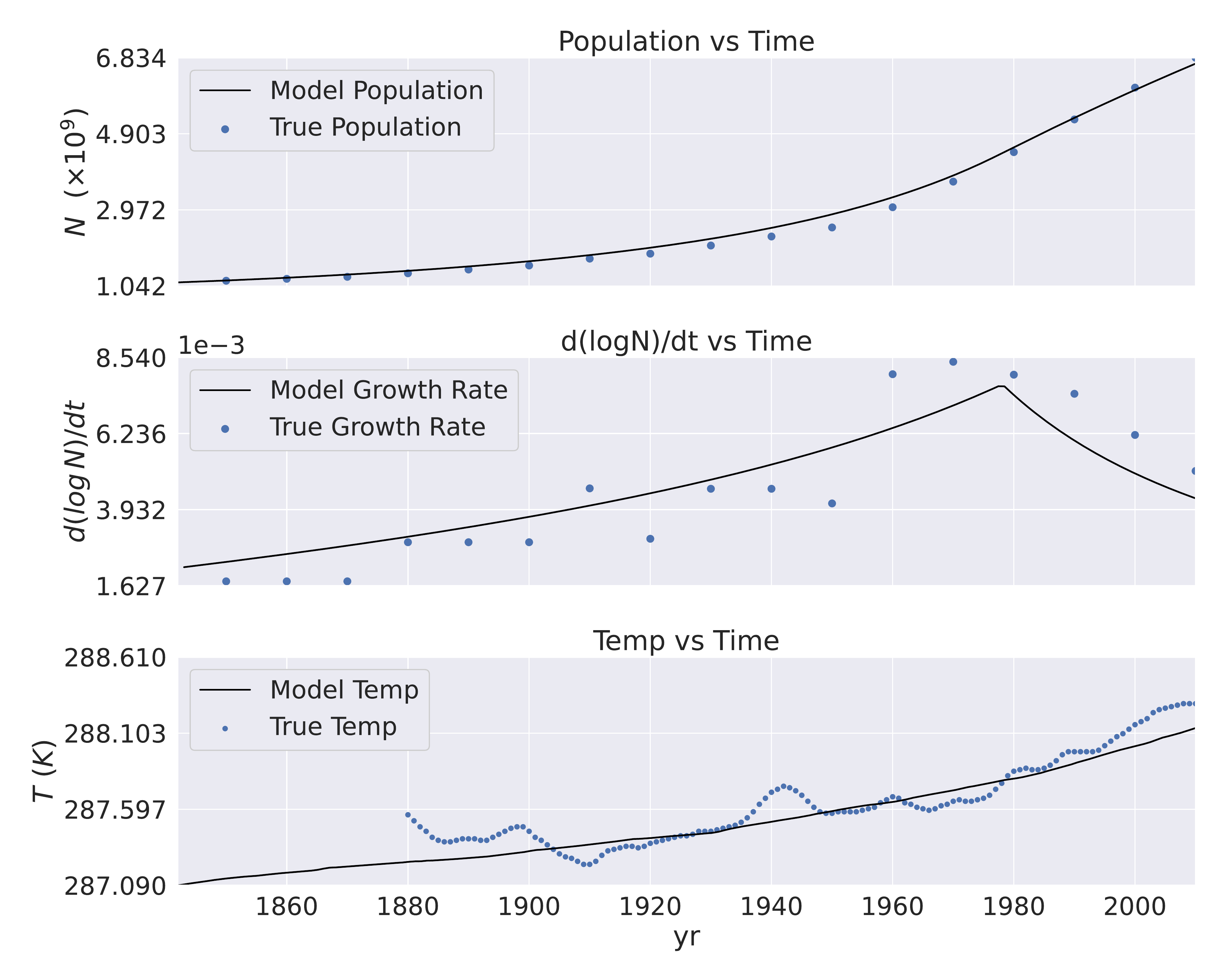}
    \caption{Model Output for Earth's recent history. By using established parameters and tuning the growth rate models we recover population (N), its first derivative and temperature (T).  Note model results are the solid black line and global data is represented by the dotted blue line.}
\label{EarthTest}
\end{figure}

The results are shown in Figure \ref{EarthTest}, which shows the evolution of population $N(t)$, growth rate, and global mean temperature $T(t)$.  As can be seen, the model does an excellent job of tracking both the rise in temperature and population during the last two centuries.  $B_0$ is the parameter we used to tune the population part of our model.  $A_0$ was fixed by our assumption that the average time between births was approximately $25$ years.  We then adjusted $B_0$ in order to have our model fit the population data we have (shown as the top-most plot of Figure \ref{EarthTest}).  Finally, we adjusted the per-capita carbon footprint $(C)$ in order to match our climates response to population growth, thus matching our global trends in temperature (shown as the bottom-most plot of Figure \ref{EarthTest}).

\section{Analytic Modeling}
Before we begin numerical integration of our equations, we first explore aspects of the solutions that can be extracted from a semi-analytic approach.  We begin by noting that its possible to approximate the climate response to increased $P=pCO_2$ via a simplified logarithmic dependence \citep{HuangBani2014}.

\begin{equation}
    T\approx T_0 + \Delta T_F \log{\frac{P}{P_0}}
    \label{eq:logresponse}
\end{equation}
\noindent
where $\Delta T_F$ is the change in temperature required for the climate sensitivity to drop by a factor of $e$ and is $\approx 4 K$ for present day Earth \citep{IPCC2014}.

Also note that with the above approximation, 

\begin{equation}
    \frac{dT}{dP} =  \frac{\Delta T_F}{P_0} e^{-\frac{T-T_0}{\Delta T_F}} \equiv D e^{-\frac{T-T_0}{\Delta T_F}}
    \label{eq:climateSensitivity}
\end{equation}

This allows us to eliminate the $CO_2$ partial pressure as an independent variable and the model reduces to

\begin{align}
\frac{dN}{dt} & = \min \left [R_{0} N \left ( 1 + \frac{P_0}{\Delta P} \left(e^{\frac{T-T_0}{\Delta T_F}}-1 \right) \right ) , R_{0} (N_{\max}-N) \right ] \notag\\
& - R_{0} N\left ( \frac{T-T_0}{\Delta T} \right)^2\\
\frac{dT}{dt} & = \, C N D e^{-\frac{T-T_0}{\Delta T_F}}
\end{align}

We can make the equations dimensionless by dividing the first equation by $R_{0} N_{\max}$ and the second by $R_{0} \Delta T$

\begin{align}
\frac{d\eta}{d\tau} &  = \min \left [ \eta \left (1 + \frac{\theta}{\alpha} \left(e^{\alpha \epsilon} - 1\right) \right), 1-\eta \right ] -\eta\epsilon^2\\
\frac{d \epsilon}{d \tau} & = \gamma \eta e^{-\alpha \epsilon}
\end{align}

The dimensionless population $(\eta)$ represents how close the civilization is to its carrying capacity.  The dimensionless temperature anomaly $(\epsilon)$ represents how much the temperature has changed from its initial value, relative to the temperature change required for the diminished growth term to counteract the benefits of the enhanced term.  Our dimensionless time $(\tau)$, is calculated with units of the population growth timescale, $t_G=1/R_0$, as discussed in Appendix A.  The other three parameters $\theta$, $\alpha$, and $\gamma$ determine the behavior of the system and are defined in Table \ref{tab:dimQs}.

In what follows we will define a climate anthropocene as the trajectory in which the population growth is strongly truncated by the increase in planetary temperature. This means the population never gets close to the natural carrying capacity of the planet.  The most important parameter with regards to whether a civilization goes through a climate anthropocene is the normalized climate forcing $\gamma$.  It represents how quickly the climate would change if the population were to reach the carrying capacity.  The parameter $\theta$ represents the expected change in population growth rate due to the consumption of fossil fuels as the temperature changes by $\Delta T$.  The parameter $\alpha$ represents the drop in the climate sensitivity (as a number of e-foldings) as the temperature changes by $\Delta T$.  One additional parameter, $\beta \equiv \gamma \theta$, is also important as it is independent of the initial climate sensitivity $(D)$ and population temperature tolerance $(\Delta T)$.  It reflects the degree to which $CO_2$ consumption increases the birth rate per natural growth time assuming the population was at the carrying capacity.

Our best fit Earth model had a $\gamma=1.94$, $\theta = 14.45$, and $\alpha=1.52$.  See Table \ref{tab:dimQs} for a summary of the dimensionless parameters.

\begin{table}[]
\caption{Dimensionless Model Quantities}
\label{tab:dimQs}
\begin{tabular}{|c|c|p{.4\columnwidth}|}
\hline
{\bf Var} & {\bf Definition}  & {\bf Description}\\ \hline
$\eta$ & $N/N_{\max}$  & Normalized population \\ \hline
$\tau$ &  $R_0 t$ & Normalized time \\ \hline
$\epsilon$ &  $(T-T_0)/\Delta T$ & Normalized temperature\\ \hline
$\theta$ & $\Delta T /(D \Delta P)$ & Normalized Birth rate acceleration\\ \hline
$\gamma$ &  $(D C N_{\max})/(R_{0}\Delta T)$ & Normalized forcing \\ \hline
$\alpha$ & $\Delta T/\Delta T_F$ & Ratio of temperature change to affect biology to temperature change to affect climate sensitivity\\ \hline
$\beta$ & $\left ( C N_{\max} \right) / \left ( R_0 \Delta P \right )$ & Increase in birth rate after burning sufficient CO2 to change temperature by $\Delta T$ \\
\hline
\end{tabular}
\end{table}

\subsection{Low $\alpha$ limit}
It is first worth considering the role played by $\alpha$ in the analysis.  The parameter $\alpha$ is the ratio of the temperature change required to affect population growth $(\Delta T)$ relative to the temperature change needed for the climate sensitivity to decrease $(\Delta T_F)$.   For present day conditions on Earth, $\Delta T_F \approx 4\, K$ \citep{IPCC2014}.  This means that if $4 \,K$ of warming is sufficient to impact the rate of human population growth $(\Delta T < 4)$, then $\alpha \lesssim 1$.  In the limit where $\alpha \ll 1$ we can take only the highest order terms and our equations become
\begin{align*}
\frac{d\eta}{d\tau} &  = \min \left [ \eta \left (1 + \theta \epsilon \right), 1-\eta \right ] -\eta\epsilon^2\\
\frac{d \epsilon}{d \tau} & = \gamma \eta \\
\end{align*}

\color{black}

\subsubsection{Low Climate Forcing $(\gamma\ll1)$}
\begin{figure}
    \centering
    \includegraphics[width=3in]{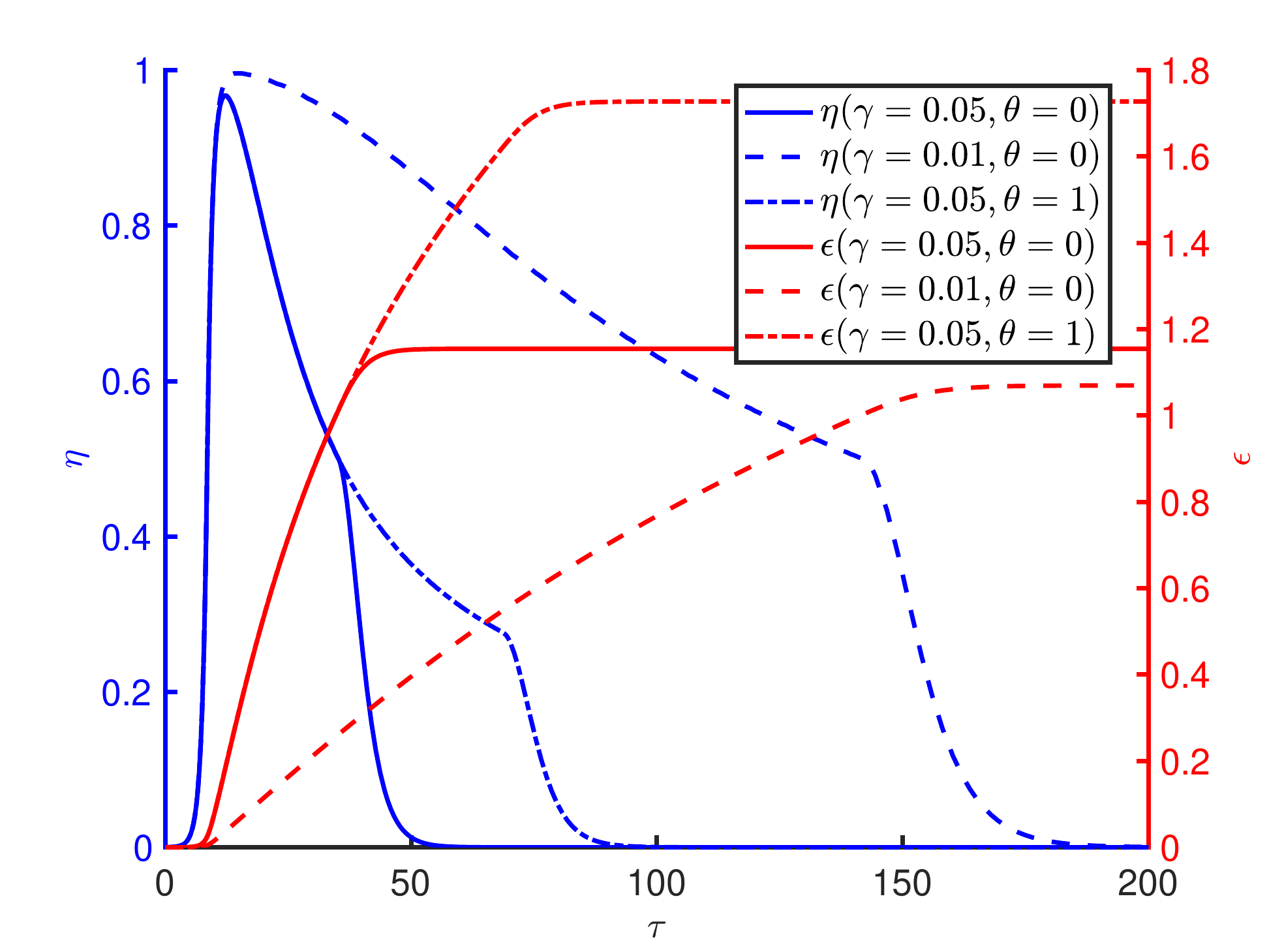}
    \caption{Evolution of population $\eta$ and temperature $\epsilon$ for cases with low forcing $\gamma << 1$ and constant climate sensitivity $\alpha = 0$.  The population reaches carrying capacity quickly followed by a slower rise in temperature over a timescale $\approx \gamma^{-1}$}
    \label{figs:lowgamma}
\end{figure}

If $\gamma$ is small, the climate forcing $d\epsilon/d\tau$ remains small even as the population reaches the carrying capacity $(\eta \rightarrow 1)$.  After the population reaches the carrying capacity, the climate will be slowly forced on a timescale $\gamma^{-1}$, causing the population to also decline on the same time scale.  Figure \ref{figs:lowgamma} shows trajectories for the semi-analytic model for $\gamma = .01$, $\gamma = .05$ and $\theta=0$ as well as a model with $\gamma=.05$ and $\theta=1$.  Note the exponential rise of the population $(\eta)$ to the carrying capacity followed by a slower, linear rise in the temperature $(\epsilon)$ over time scales $\approx \gamma^{-1}$.  The change in slope during the decline is due to our birth rate (regardless of population and technology) being capped at $1-\eta$ as a means of implementing a carrying capacity.  As the population begins to decline, for a while it is able to maintain the maximum birth rate (which itself is increasing).  This helps to counter the increased death rate due to the changing climate.  Eventually, however, the population drops far enough that the technology enhanced birth rate is less than the enhanced death rate at which point the population drops off more precipitously.  For $\theta=0$ this occurs at $\eta=.5$ while for the case with $\theta=1$ this transition occurs later as technology is able to assist in increasing the birth rate for a longer time.  Also note in this case that $\theta$ has little affect on the initial growth or early decline.  For this model $\beta \equiv \gamma \theta \ll 1$ so $CO_2$ consumption is not significantly altering the growth rate until after the population peaks.  Also,  $\theta$ does not affect the initial decline because the birth rate is at the carrying capacity limit, $1-\eta$, and is independent of the amount of $CO_2$ consumed.

\subsubsection{High Climate Forcing $(\gamma \gg 1)$}
In the limit of high climate forcing, the population is able to force (i.e. change) the climate while being well below the carrying capacity ($\eta < 1$).  The population will then grow exponentially (or faster if aided by technology when $\theta \gtrsim 1$).  The temperature anomaly $(\epsilon)$ will also grow exponentially (or faster) until $\eta \gtrapprox \frac{1}{\gamma}$ and $\epsilon \gtrapprox 1$.  

\begin{figure}
    \centering
    \includegraphics[width=3in]{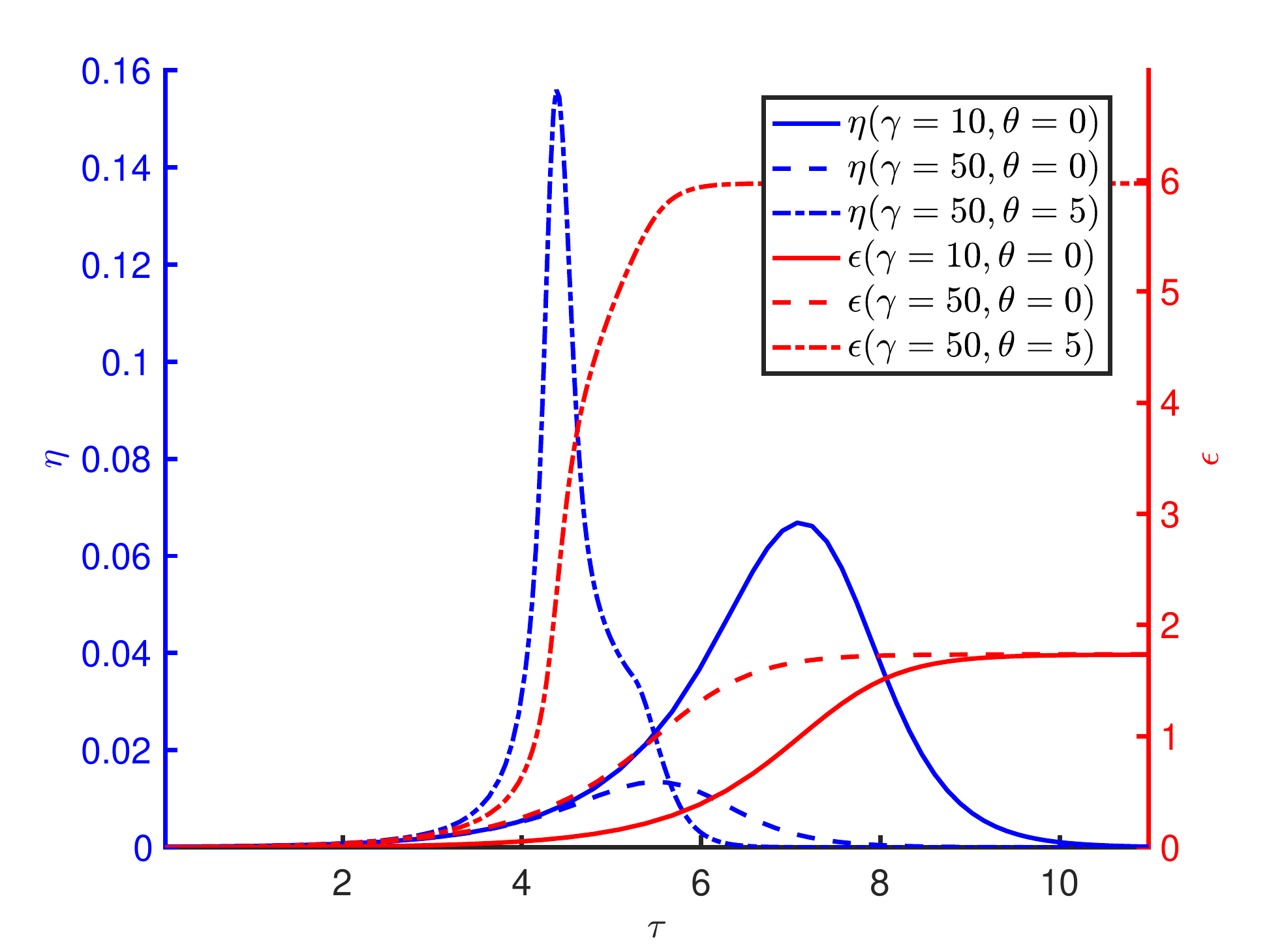}
    \caption{Evolution of population $\eta$ and temperature $\epsilon$ for cases with high forcing $\gamma \gg 1$ and constant climate sensitivity $\alpha=0$.  The population pushes the climate well before reaching the carrying capacity followed by a faster collapse on the order of the growth time - or even faster if the growth rate going into the Anthropocene is accelerated by technology.}
    \label{figs:highgamma}
\end{figure}

Figure \ref{figs:highgamma} shows trajectories for cases with $\theta=0$ and $\gamma = 10$ and $50$.  In both cases the population rises exponentially and then begins to turn over when $\epsilon \approx 1$ and $\eta \approx \frac{1}{\gamma}$.  The case with the larger $\gamma=50$ forces the temperature $\epsilon$ faster and as a result has a smaller peak population which it achieves earlier.  In both cases the timescale for the population to decline is of order $\tau = 1$.  

Figure \ref{figs:highgamma} also shows the trajectory for the case with $\gamma=50$ but with a technology benefit $\theta = 5$.  In that case the rise is accelerated due to a technology assisted enhanced birth rate resulting in not only an earlier peak, but one that also has a higher population. 
This  results in faster environmental forcing which then leads to a shorter collapse time. 
For very large $\theta$, the min function acts to ensure that technology does not accelerate birth rates to an unrealistic value by enforcing a maximum growth rate given by $1-\eta$.  Thus, in these cases, initially $\theta$ accelerates births so much that the maximum rate is exceeded, at which point this maximum rate takes over, allowing the civilization to smoothly enter their collapse.  As they begin to collapse, $\eta$ begins to decrease so that $1-\eta$ starts to increase.  Eventually we return to the initial state where $1-\eta>\eta(1+\theta\epsilon)$.  At this point the min function switches to using the technology enhanced birth term, which has the effect of softening the decline.  This can be seen during the collapse of the trajectory with $\gamma=50\ \&\ \theta=5$ in Figure $\ref{figs:highgamma}$ as a slight deviation in the slope of $\eta$ (i.e. a 'shoulder').

Now we can estimate the collapse time by looking at the second derivative at the peak population.

\begin{align*} 
    \tauColl & = \left . \sqrt{-\frac{\eta}{ \frac{d^2\eta}{d \tau^2}}}\right |_{\frac{d\eta}{d\tau}=0} \\
\end{align*}

To simplify things we will assume that $\gamma$ is large enough so that the technology assisted growth rate never exceeds the carrying capacity limit $(1-\eta)$.  In that case we find that (see Appendix B)

\begin{equation}
    \tauColl=\frac{1}{\max(\sqrt{2},\ \theta)}\quad  (\gamma\gg1)
\end{equation}

\noindent
Combining this with the decline times for the $\gamma \ll 1$ case, we have
\begin{equation}
\label{eq:tcoll}
\tauColl = \max \left [\frac{1}{\gamma},\ \frac{1}{\max(\sqrt{2},\theta)} \right]
\end{equation}

\subsection{Role of $\alpha$}
\begin{figure}
    \centering
    \includegraphics[width=3in]{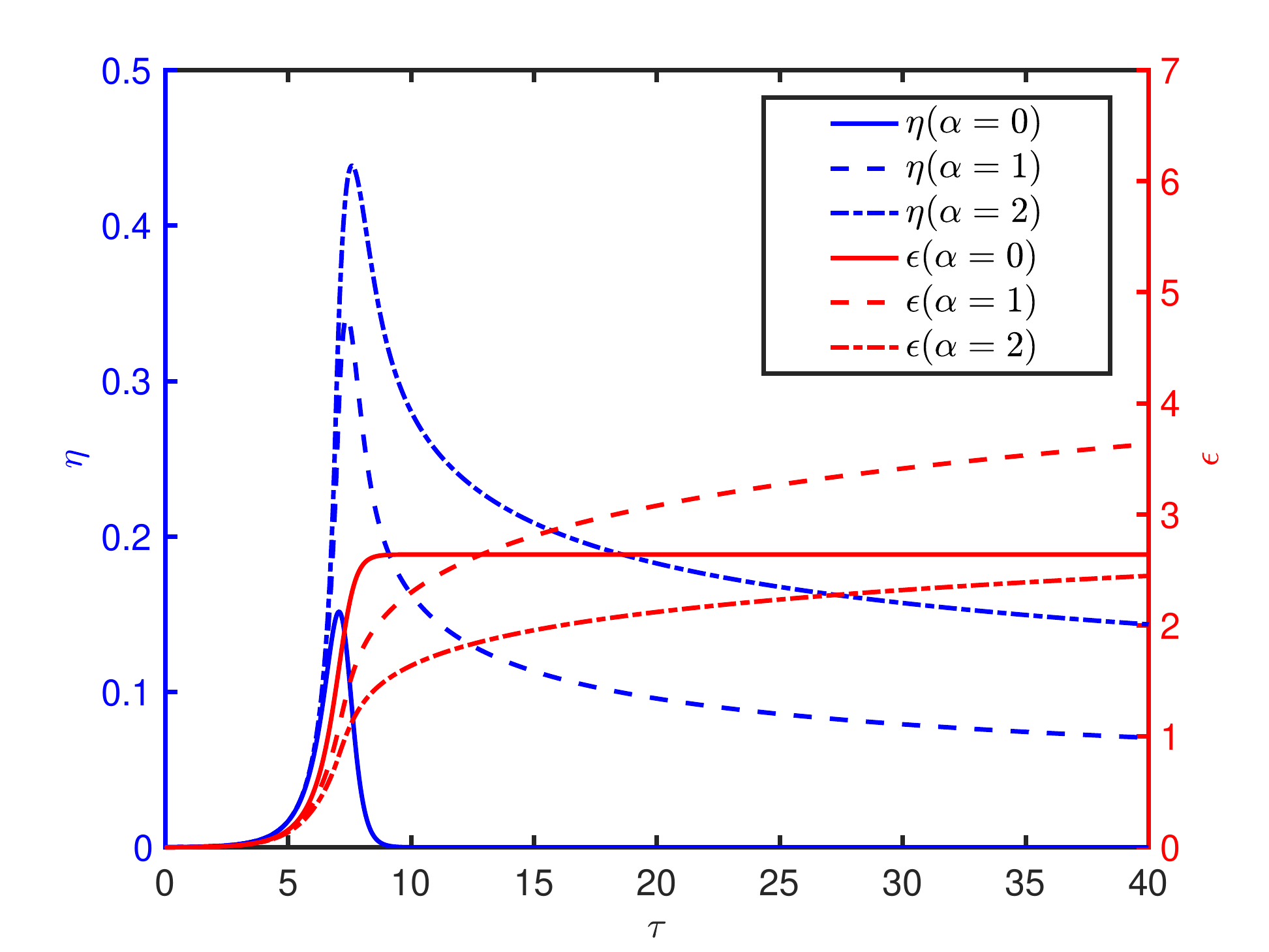}
    \caption{Evolution of population $\eta$ and temperature $\epsilon$ for cases with high forcing $\gamma = 10$, a modest CO2 birth benefit $\theta = 1$ and different values for $\alpha$.  Cases with larger $\alpha$ have shallower declines because the climate sensitivity is decreasing as the temperature increases.}
    \label{figs:alpha}
\end{figure}

  Finally we turn to the role of $\alpha$, which determines how much the climate sensitivity will decrease as the temperature changes.  Figure \ref{figs:alpha} shows the results of runs with $\gamma=10$, $\theta=1$, and $\alpha = 0, 1,\ \mbox{\&}\  2$.  In the cases with $\alpha > 0$, increasing temperatures will reduce the climate sensitivity before reducing net growth rates, allowing civilizations to reach higher peak populations while also delaying the time for them to be reached.  Furthermore, the initial decline begins to taper off as both the decreasing climate sensitivity and increasing birth rate act to reduce the decline in population.  It may be unreasonable to expect the birth rate to continually increase over many generations due to $CO_2$ consumption, however $CO_2$ consumption might also act to mitigate the amount of increased death due to changing temperatures, resulting in a similar tendency.
  
\section{Results of Fully Coupled Model}
It is important to understand the meaning of the solution domains we have just explicated in terms of the goal of the study.  We are interested in the ubiquity of climate dominated anthropocenes.  Thus in this study we ask, given different planetary initial conditions, how often will a civilization's energy harvesting (in this case via combustion) lead to population growth which then leads to rapid climate change which, finally, leads to rapid population {\it decline}.  Addressing this question is the specific intent of our modeling.  As demonstrated by the analysis above, however, some solutions exist in which the population rises to the planet's carrying capacity {\it before} the changing climate produces adverse effects. This would not be a climate dominated anthropocene in our definition.  

\begin{figure}
    \centering
    \includegraphics[width=\columnwidth]{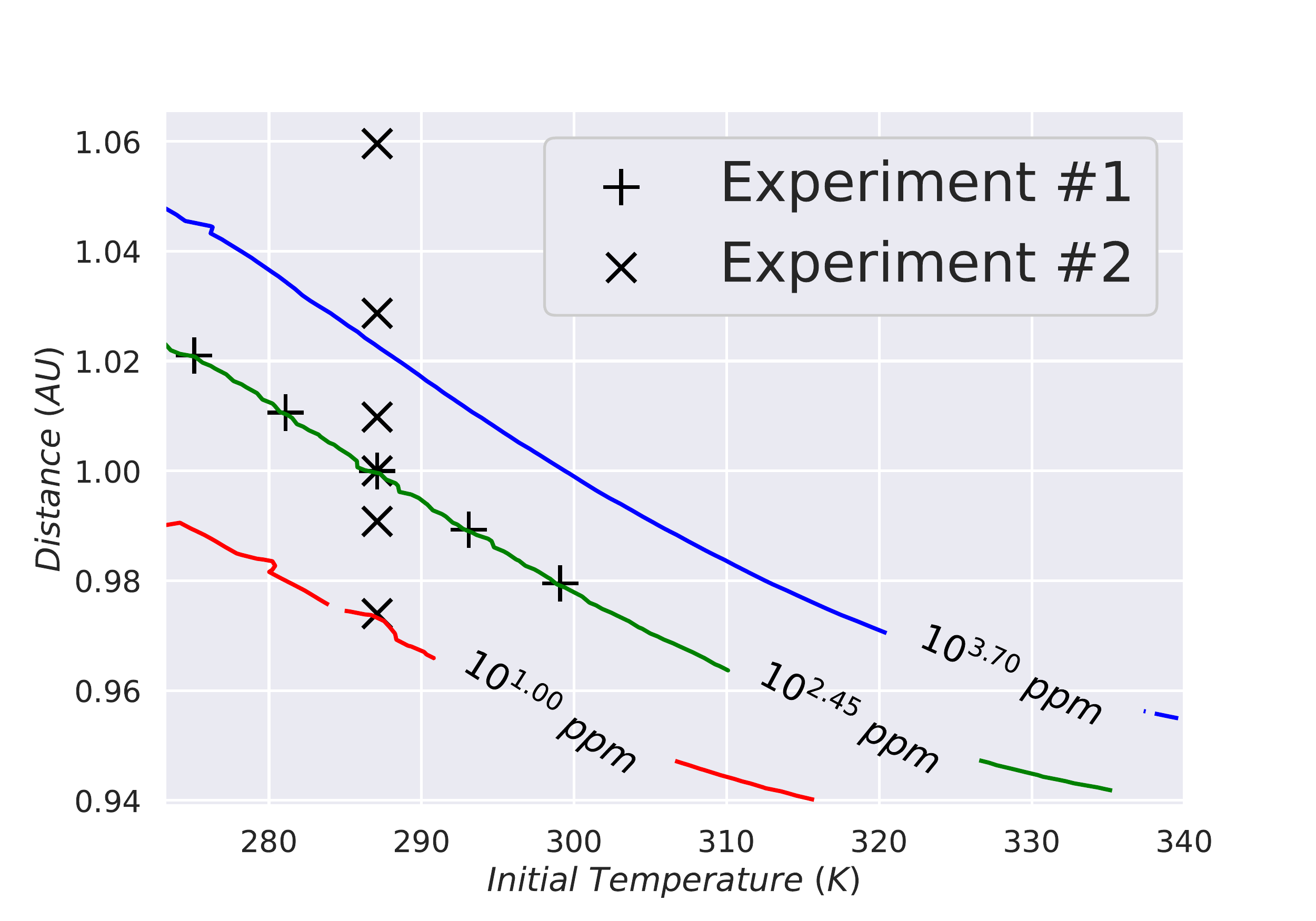}
    \caption{Parameter space and model locations for two initial experimental sets. Experiment \#1: Constant Composition was defined by constant levels of initial $pCO_2=284\ ppm$, and is denoted by the pluses ($+$).  Experiment \#2: Constant Temperature was defined by constant initial temperature $T_{0}=287.09K$, and is denoted by the crosses ($\times$).  The two experiments intersect at Earth when initial $pCO_2=284\ ppm$ and $T_{0}=287.09$.}
    \label{figs:exps}
\end{figure}

It is, however, worth noting that a rapid (exponential) population rise to the host world's carrying capacity $(N_{max})$ will bring its own potentially existential challenges.  The very definition of carrying capacity implies that at when $N = N_{max}$, the civilization is at the edge of what the planet can provide in terms of "ecosystem services".  Thus, while these classes of systems will not fall under our definition of climate-dominated anthropocenes, they should not be considered to be cases that have escaped the possibility of rapid population declines or even collapse.  It is simply that our models do not include the processes (i.e. biospheric feedbacks) which could produce them.

\subsection{Results: Constant Temperature and Composition Models.}
We now return to the full non-linear model described in section 2.  We have carried forward a large suite of numerical experiments using this model with the goal of investigating how the trajectory of coupled planet-civilization systems depend on various initial conditions.  To recap we set up our initial conditions with two key assumptions. (1) The biology of the organisms building the civilization requires liquid water, so the host planet must be within the star's habitable zone. (2) The organisms have temperature and $pCO_2$ limits beyond which they can not survive.  Taken together, these two conditions define a "Complex Life Habitable Zone" (CLHZ \cite{Schwietermanea2019}), as discussed in the introduction. For illustrative purposes, we will begin our study by using values similar to Earth and human life for these limits, but will always consider them to be free parameters.

\begin{figure*}
    \centering
    \includegraphics[width=3.5in]{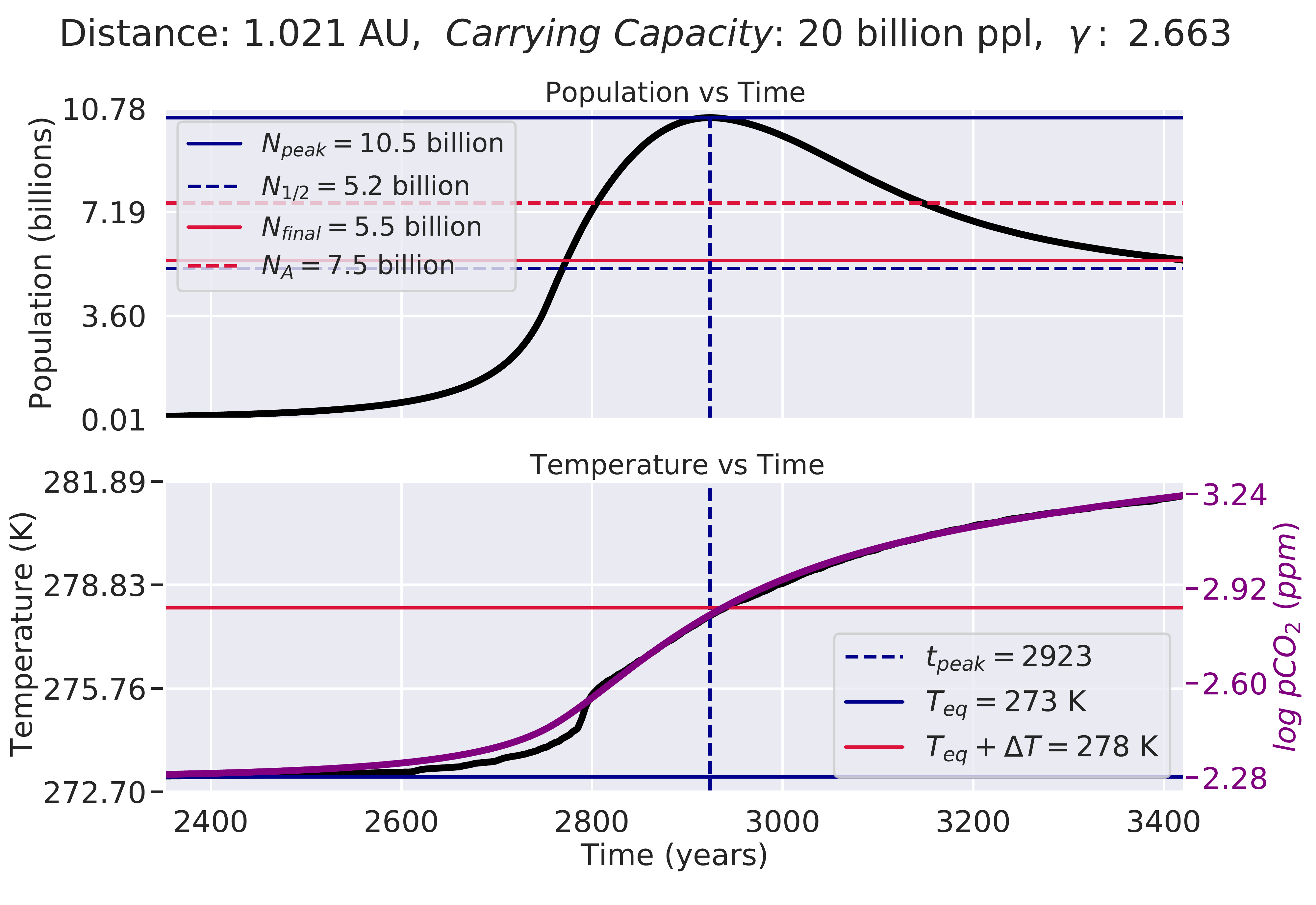}
    \includegraphics[width=3.5in]{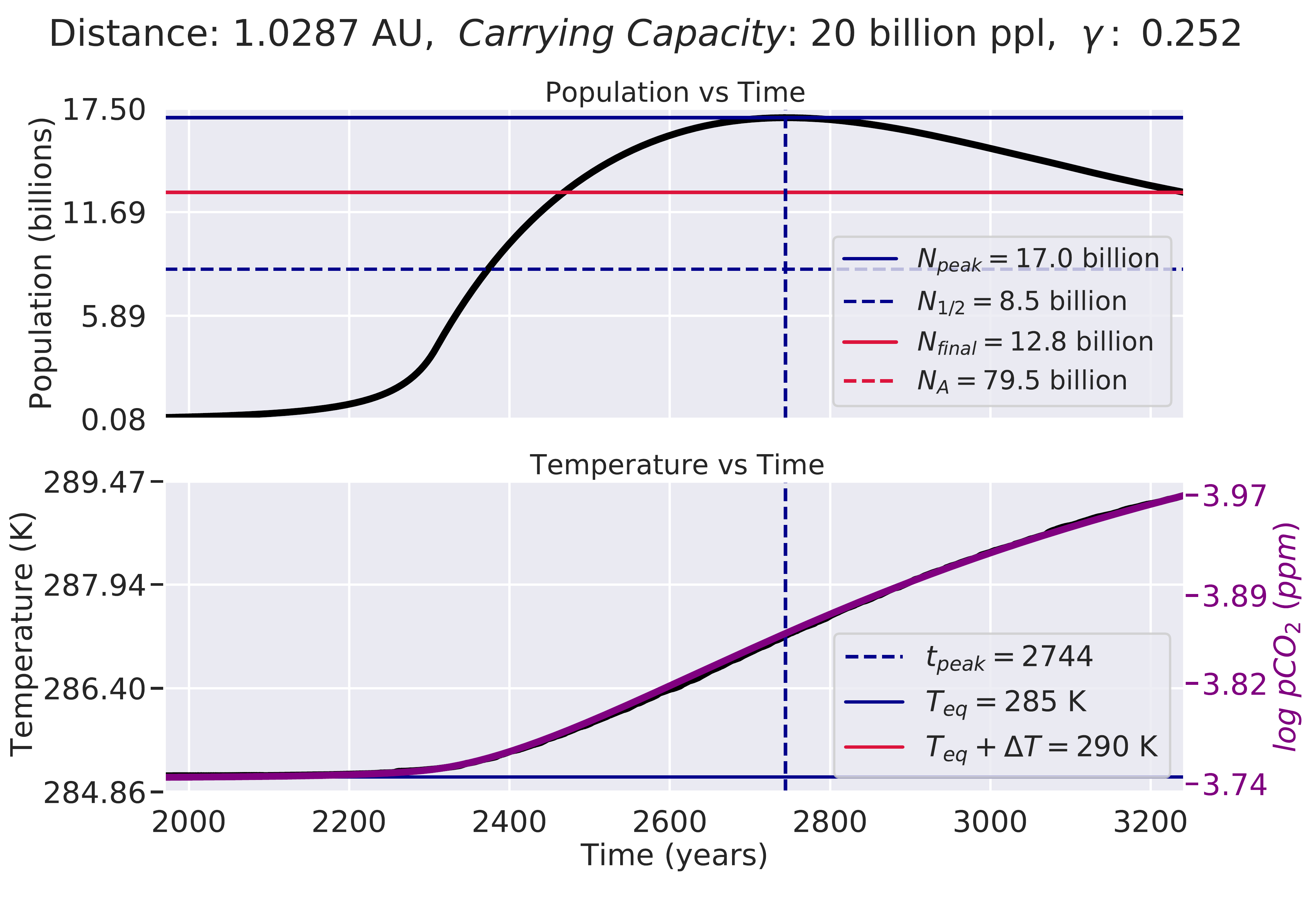}
    \includegraphics[width=3.5in]{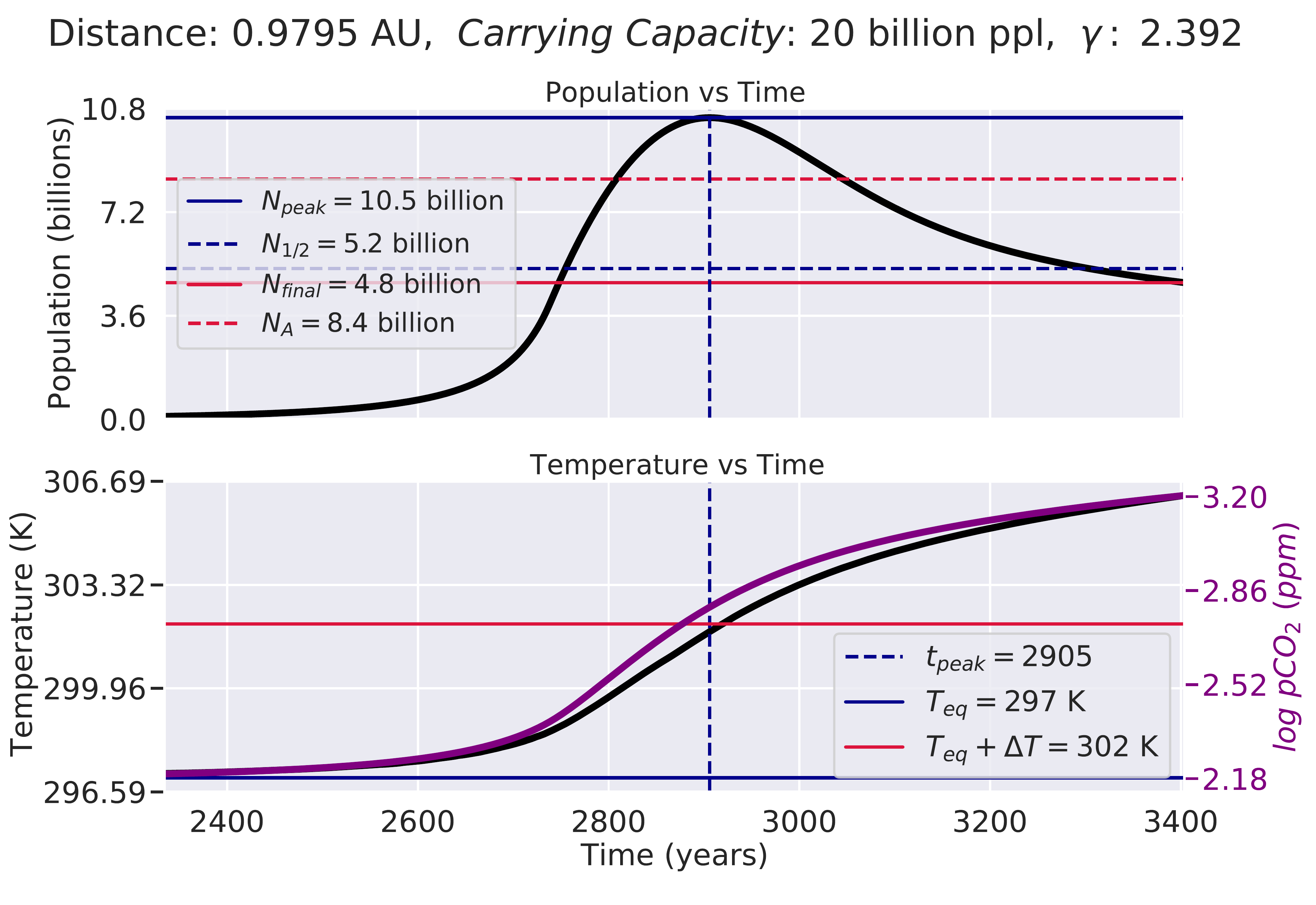}
    \includegraphics[width=3.5in]{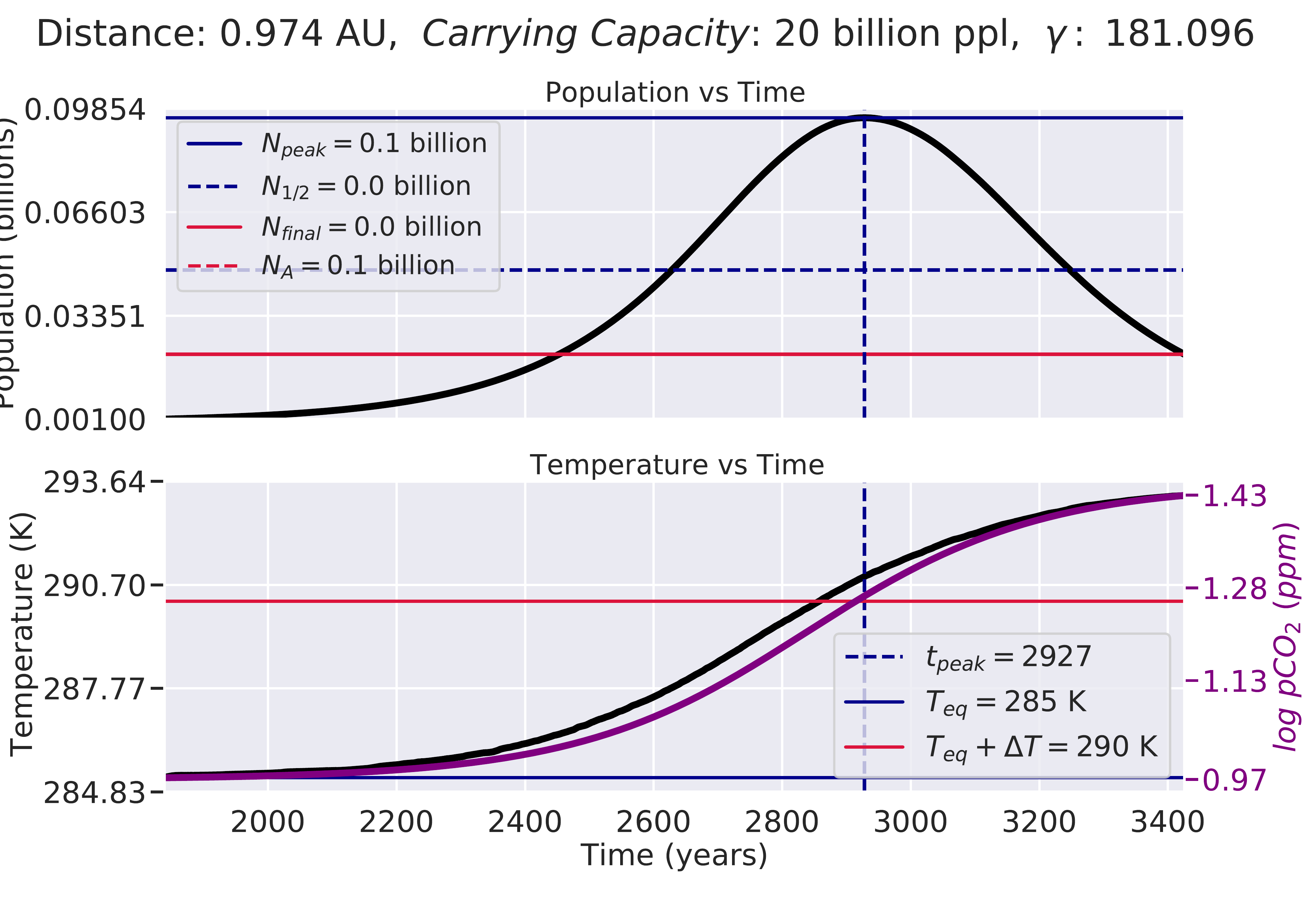}
    \caption{Selected trajectories.  The left column shows the highest and lowest distance from experiment \#1 (top and bottom respectively).  Similarly, the right column shows the same for experiment \#2.  Note that on the left column, the value of $\gamma$ changes marginally, this is due to the marginal difference between values of $dTdP$ at these two locations.  In contrast, on the right column, the change in gamma is drastic, corresponding to the drastic change in $dTdP$ resulting from different levels of initial $pCO_2$.}
    \label{trajectories}
\end{figure*}

We focus on two initial planetary parameters: the orbital distance from the planet to its host star $(a)$, and the initial chemical composition of the atmosphere quantified in terms of $P=pCO_2$.  The effect of these parameters on the models are not fully independent.  Planets on the inner edge of the habitable zone ($a<1$) have climates that are less stable against small increases in $CO_2$ when compared to similar perturbations on dense $CO_2$ planets near the outer edge ($a>1$).  We chose our boundaries in $pCO_2$ to have an outer edge corresponding to $5,000\ ppm$ as this represents the upper limit of $pCO_2$ amenable for animal life on Earth \citep{WittmannPortner2013}.  For a lower value we choose  $10\ ppm$ (we could not go to zero due to limitations imposed by our EBM).


We begin with two sets of experiments that illustrate the basic behavior of the civilization-planet system. The first of these are "constant composition" models.  These keep initial $pCO_2$ constant and allow the initial (equilibrium) planetary temperature, $T_0$, to vary as we change the orbital distance $(a)$.  Using $T_0 =287.1 K$ as our fiducial value, we ran four additional models with temperatures  evenly spaced above and below $T_0$ in steps of $6K$.  This spacing was chosen in order to have all models safely within the habitable zone ($273K<T_0<373K$).  In Figure \ref{figs:exps} we show the location of the models in the ($a, T_0$) plane.  This representation is important because we will later overlay contours of various quantities such as $\gamma$ and the collapse time $\tauColl$  when we run a larger array of models that sweep across ($a, T_0$) space.

The second set of experiments are "constant temperature" models.  These keep initial temperatures $(T_0)$ constant and allow the initial $pCO_2$ to vary as we change the orbital distance $(a)$.  Five models were run centered on Earth's current $pCO_2$ concentration with two below and three above, each spaced by $log(pCO_2)=0.7$.  The highest, $P_0 = pCO_{2,0} \approx  28,000\,ppm$ was meant to illustrate the evolution a system with $CO_2$ concentration beyond what animal life on Earth can tolerate.

We now explore the trajectories occurring along our lines of constant $T_0$ and $a$.  Figure \ref{trajectories} shows four runs from our two sets of experiments. Consider first the model in the upper left of the figure which corresponds to a constant initial composition case with $a=1.021$ AU.  This model begins with $log(pCO_2)_0 = 2.28$. Using the climate model to determine the planet's climate sensitivity $dT/dP$, along with the other model parameters, yields a $\gamma = 2.663$. Since $\gamma >1$ we expect this model to experience a climate anthropocene.  This is, in fact, what occurs as we see a steep rise in population beginning at approximately $t=2700$ y. The population then peaks at $N_{peak} = 10.5$ billion individuals about two centuries later. After the peak, the population declines by half over the following two centuries. The cause of the peak and decline can be see in the rising $pCO_2$ levels. At $N_{peak}$ we find $log(pCO_2)_0 \sim 2.9$. The increased planetary greenhouse effect has driven temperatures almost past the civilization's tolerance $\Delta T$.  Note that in this model the population never comes close to the planet's carrying capacity of $N_{max}= 20$ billion. 

In the lower left panel of Figure \ref{trajectories} we show a second constant initial composition case, though in this model $a=0.9795$ AU.  Since $log(pCO_{2,0})$ is the same as the model above, it also has $\gamma = 2.392$ and we expect a climate anthropocene. Because the $\gamma$ values for these two runs are relatively similar we find similar trajectories in terms the rise, peak and decline in population $N$ along with a relatively steady increase in $log(pCO_2)$ and $T$.  

In the upper right panels of Figure \ref{trajectories} we show a constant $T_0$ model with $a=1.029$ AU.  This world is further from its (solar type) star so it requires a higher $pCO_2$ to maintain $T_0 = 287$ K.  Thus with $log(pCO_2)_0 = 3.74$ we find a $\gamma = 0.252$.  In this case $\gamma<1$ and we expect this model to avoid a climate anthropocene. The results show this to almost be the case.  The population peaks at $N_{peak}/N_{max} = .89$.  Thus the initial exponential growth phase of the population is able to carry the civilization close to the planets' carrying capacity before a decline sets in.  Note also the longer period between the onset of initial exponential growth and time for the population to reach $N_{peak}$.

Finally in the lower right we show the trajectory for second case with a constant initial temperature.  This planet is closer to its sun with $a=.974$ AU and, as such, a significantly lower initial $pCO_2$ is required ($\log(pCO_{2,0}) = 0.97)$.  Because this world is closer to it's star and begins with a lower $CO_2$ concentration it is more sensitive to changes in $pCO_2$. This is reflected in its value of $\gamma = 181$. Examining the trajectory we once again see the rapid rise and fall of the population.  In this model however the extreme climate sensitivity which drives $\gamma \gg 1$ manifests itself in $N_{peak} \sim 10^{-3} N_{max}$.  The population never comes close to the planet's carry capacity before the climate is driven into detrimental states ($T \sim T_0 + \Delta T$).  Note that this behavior was captured in the value of the anthropogenic population defined in Appendix A, which, for this model, takes on the value $N_A = 0.1$ billion individuals.


\begin{figure}[h!]
    \centering
    \includegraphics[width=\columnwidth]{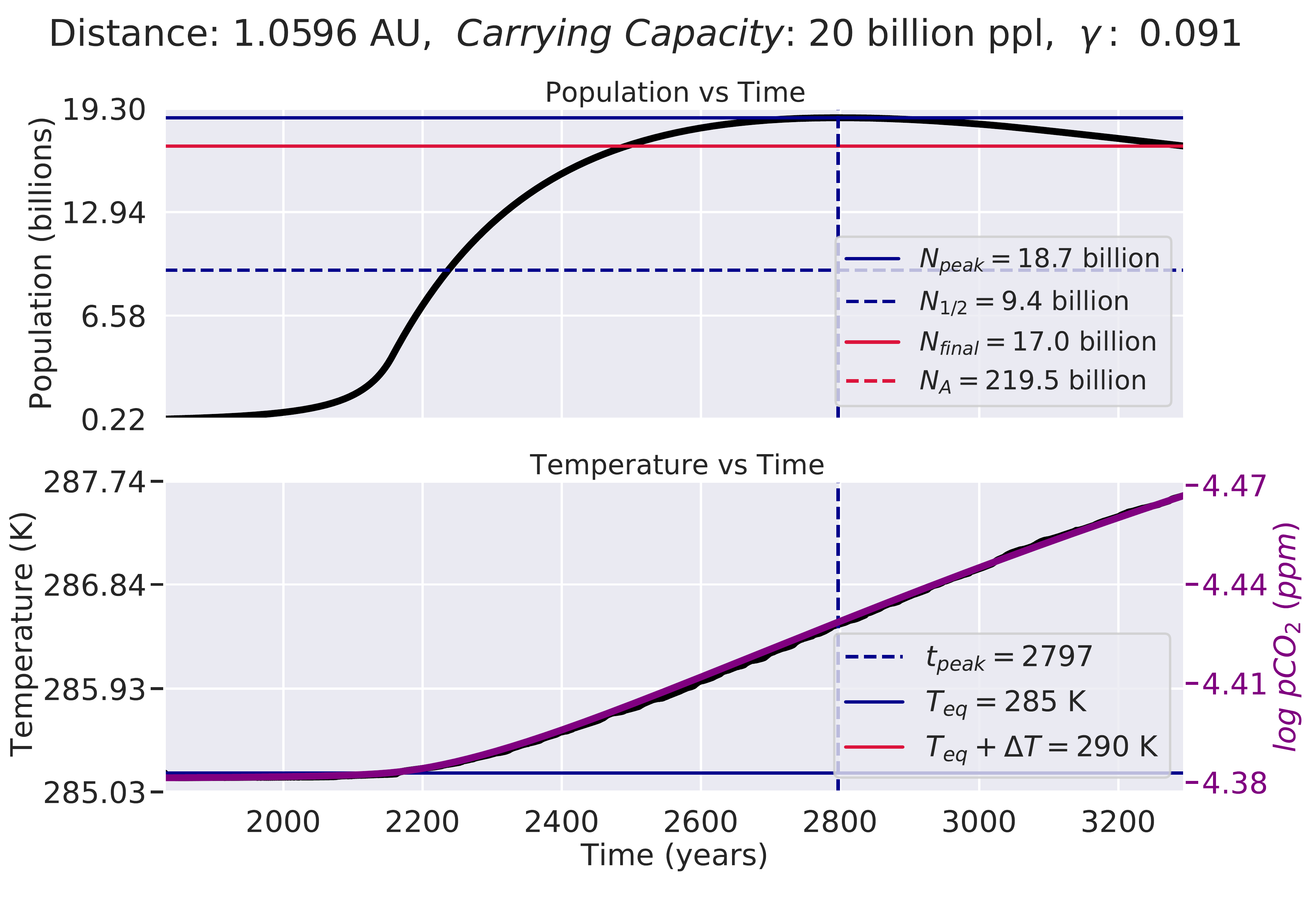}
    \caption{This model results in the civilization reaching their carrying capacity, thus overpopulating their planet.  The initial $pCO_2$ is $10^{4.38}\approx 24,000\ ppm$, which puts it in the "danger zone" for habitability. }\label{dangerZone}
\end{figure}

It is also of interest to consider the trajectories for planets with initial $pCO_2$ greater than that which Earth life can tolerate.  These worlds would orbit at larger radii $(a)$ and would require higher greenhouse gas concentrations to maintain habitable environments.  Figure \ref{dangerZone} shows such a model with $a=1.06$ AU and $\log(pCO_{2,0}) = 4.38$, yielding a $\gamma = 0.09$.  The trajectory for this model shows the population climbing to within $1.5\%$ of the planet's carrying capacity or $(N\sim N_{max})$.  When this peak occurs the global mean temperature is still well below the threshold for a climate dominated Anthropocene ($T<T_0+\Delta T$). Thus for this case the population grows until it reaches $N_{max}$ without driving the climate into a significant new state that is detrimental to the functioning of the civilization. The trajectory shown in this figure is typical for worlds with very high $log(pCO_2)_0$. The addition of more $CO_2$ via combustion does not alter the climate significantly before the population rises to levels beyond what the planet, via its carrying capacity, can sustain.

\begin{figure*}
    \includegraphics[width=7in]{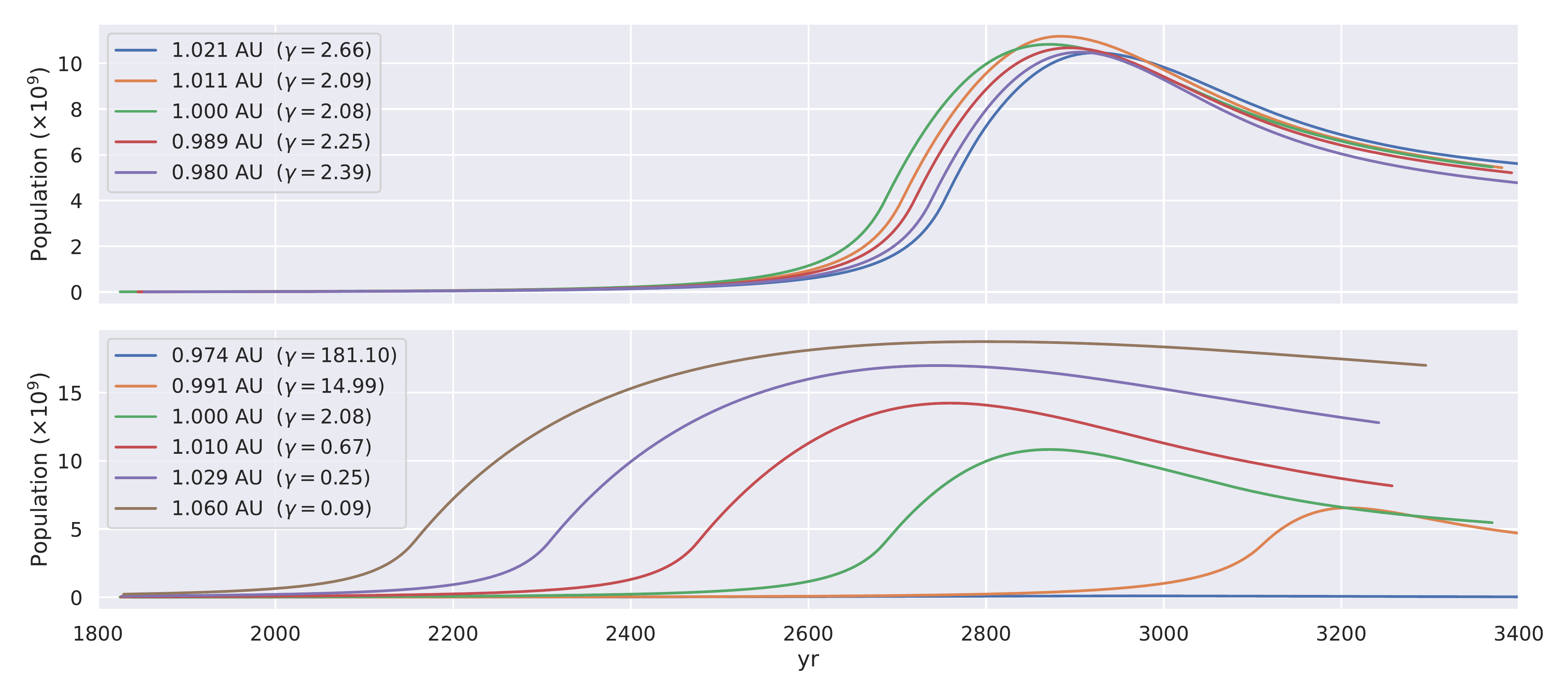}
    \caption{The top plot shows the trajectories for Experiment \#1: Constant Composition, defined by constant levels of initial $pCO_2=284\ ppm$.  The bottom plot shows the trajectories for Experiment \#2, defined by constant initial global temperatures ($T_0=287.09K$).  The most influential quantity for the co-evolution of intelligent civilizations and their planets is the quantity $dTdP$, which is a function of $pCO_2$.  The value of $pCO_2$ was found in order to make any given distance have any given temperature.  Thus, the value of initial $pCO_2$ principally determines the resulting co-evolution.  This is why the trajectories for experiment \#1 are very similar, as they all have approximately equal initial levels of $pCO_2$.  In contrast, the trajectories in the bottom plot all have the same temperature, but different distance, thus also have different levels of $pCO_2$ resulting in very different values for $dTdP$.  This is why the trajectories in this plot are much more diverse.}
    \label{fig:expsTaj}
\end{figure*}

In Figure \ref{fig:expsTaj} we show all trajectories for both experiments.  The upper panel shows results from constant initial composition models while the lower panel shows those for constant initial temperature models.  The similarities and differences both within and between the models offers insight into the dynamics and its relationship to our dimensionless parameters.  For example, the value of $\gamma$ can be seen to uniquely determine the resulting evolution of population.  Furthermore, the upper panel shows that models with constant $pCO_2$ yet different orbital radii and initial temperatures will share the same $\gamma$.

\subsection{Results: 2D Parameter Sweeps}

\begin{figure*}
    \centering
    \includegraphics[width=\textwidth]{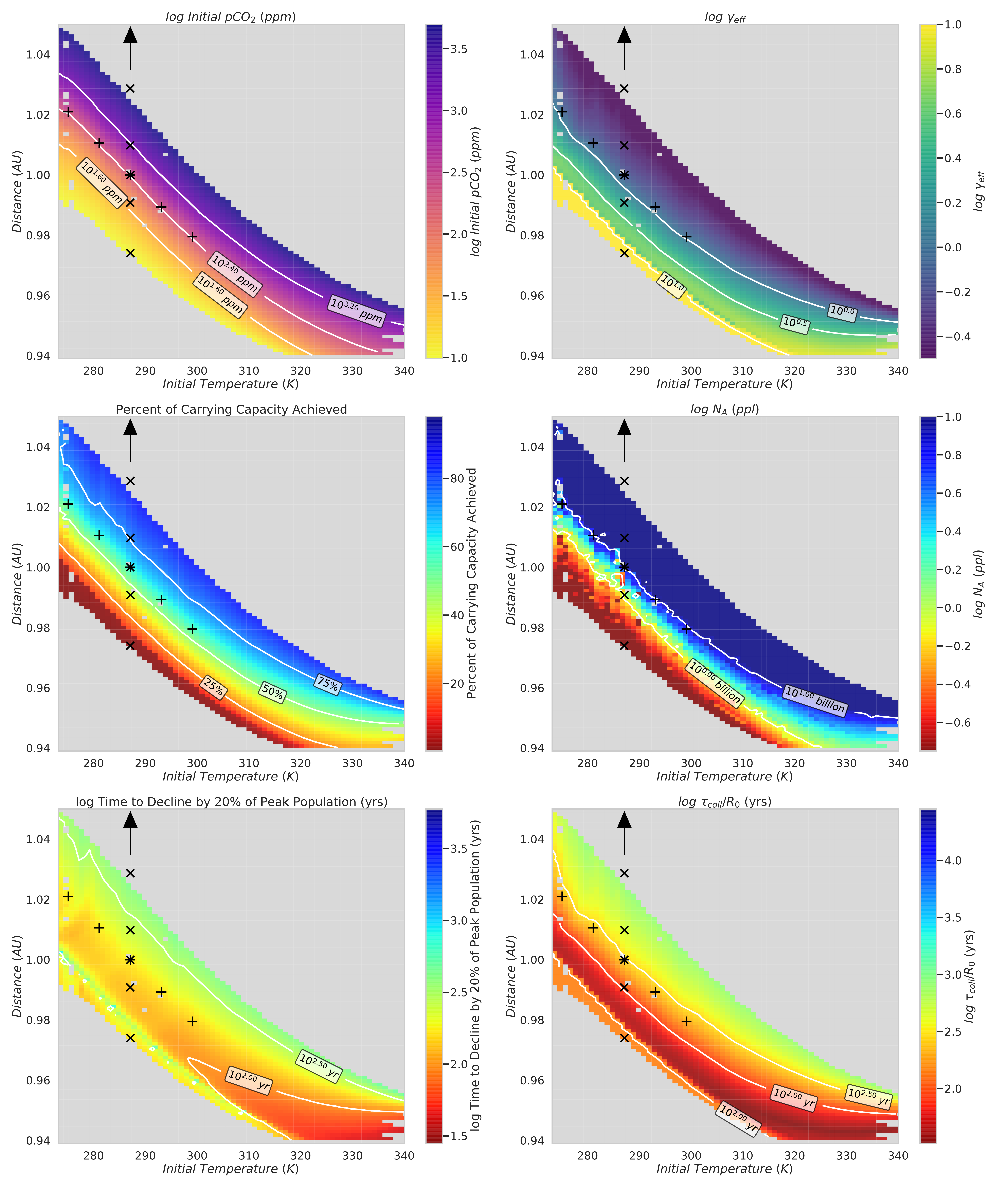}
    \caption{The plots on the left hand side show numerically calculated quantities, while the plots on the right hand side show analytically derived quantities.  The top row shows values calculated with the uncoupled energy balance model.  The middle row shows quantities related to the populations growth.  The bottom row shows timescales related to the populations collapse.  The bottom left of the plots are grey because of limitations imposed by the EBM.  The top right part of the plots are grey because the value of $pCO_2$ required there was greater than $5,000\ ppm$, a level deemed uninhabitable for long-term habitability by intelligent civilizations \citep{Schwietermanea2019}.  The black arrow points to the location of the experiment that we ran in the 'danger zone'. }
    \label{fig:paramSweepdT5}
\end{figure*}

To explore the broad dependence on initial conditions we next choose 100 different distances and temperatures for the models.  The results of this parameter sweep are shown as contour plots in Figure $\ref{fig:paramSweepdT5}$. The left column of the plots show quantities taken from the full numerical models, while the right column shows the corresponding analytical quantities. 

In the top left we present contours and color mapping of the initial atmospheric composition, $\log(pCO_{2,0})$, for all the runs.  This was calculated using only the uncoupled energy balance model. Note that we exclude models with $\log(pCO_{2,0}) > 3.7$ as being outside the CLHZ \citep{Schwietermanea2019}.

The top right panel in Figure $\ref{fig:paramSweepdT5}$ presents initial $\gamma$,  defined in Table \ref{tab:dimQs}.  Recall that $\gamma$ is dependent principally on $dT/dP|_{P=P_0}$, which itself is dependent principally on initial $pCO_2$. Thus the contour lines of initial $pCO_2$ also correspond to contour lines for $\gamma$. This plot shows that only the outermost layers of orbits have $\gamma < 1$, indicative of worlds not at risk of a climate Anthropocene.  Recall that even models with with $\gamma$ slightly less than 1 can still have their growth truncated by climate effects.  Thus we find that most civilizations arising in CLHZ will be susceptible to having their growth truncated by rising temperatures and changing climate.  

The middle row describes the population dynamics for the civilizations and focuses on parameters associated with growth.  The left column presents the numerically measured percentage of the carrying capacity each civilization reached $(N_{peak}/N_{max})$.  The right column shows the analytic predictions of the quantity $N_A$, defined in Appendix A as the number of people required to force the climate out of equilibrium in a single growth timescale. In $N_{peak}/N_{max}$ we see only the outermost orbits at each initial temperature are able to rise to their carrying capacity before increasing temperatures significantly increase death rates and halt population growth.  Note all models in the parameter sweep began with a carrying capacity of $N_{max} = 20$ billion.  Consideration of the $N_A$ contour plots demonstrates how the linear model accurately identifies the peak population possible in climate Anthropocene worlds.

Finally, the last row of plots considers what happens after $N=N_{peak}$.  On the left we show the time for population to decline by $20\%$. Here we see most of the models experience a decline on timescales of a few centuries at most, while initially hotter worlds on inner orbits can have declines over just decades.  The collapse time parameter $\tau_{coll}$ is shown on the lower right.  Once again we see timescales of order decades to a few centuries associated with significant population decline.  Note however that $\tau_{coll}$ shows a "valley" feature at intermediate orbital distances where $\tau_{coll}$ falls and then rises again as one moves outward in orbital distance along a line of constant $T_0$. We return to this valley which reflects the dependence of $\tau_{coll}$ on $1/\gamma$ in the next section.

\subsection{Dependence of Civilization Temperature Tolerance}
All the models discussed in the last section used a constant value of the civilization's temperature tolerance, $\Delta T = 5K$.  As discussed earlier this parameter is intended to capture biological, ecological and cultural effects of the civilization's response to its climate forcing.  Depending on the fragility of both the social organization and ecosystems on which it depends, a smaller $\Delta T$ may be enough to trigger detrimental consequences for the civilization. Thus, in this section we vary $\Delta T$ to asses its role in model outcomes. 

\begin{figure*}
    \centering
    \includegraphics[scale=0.475]{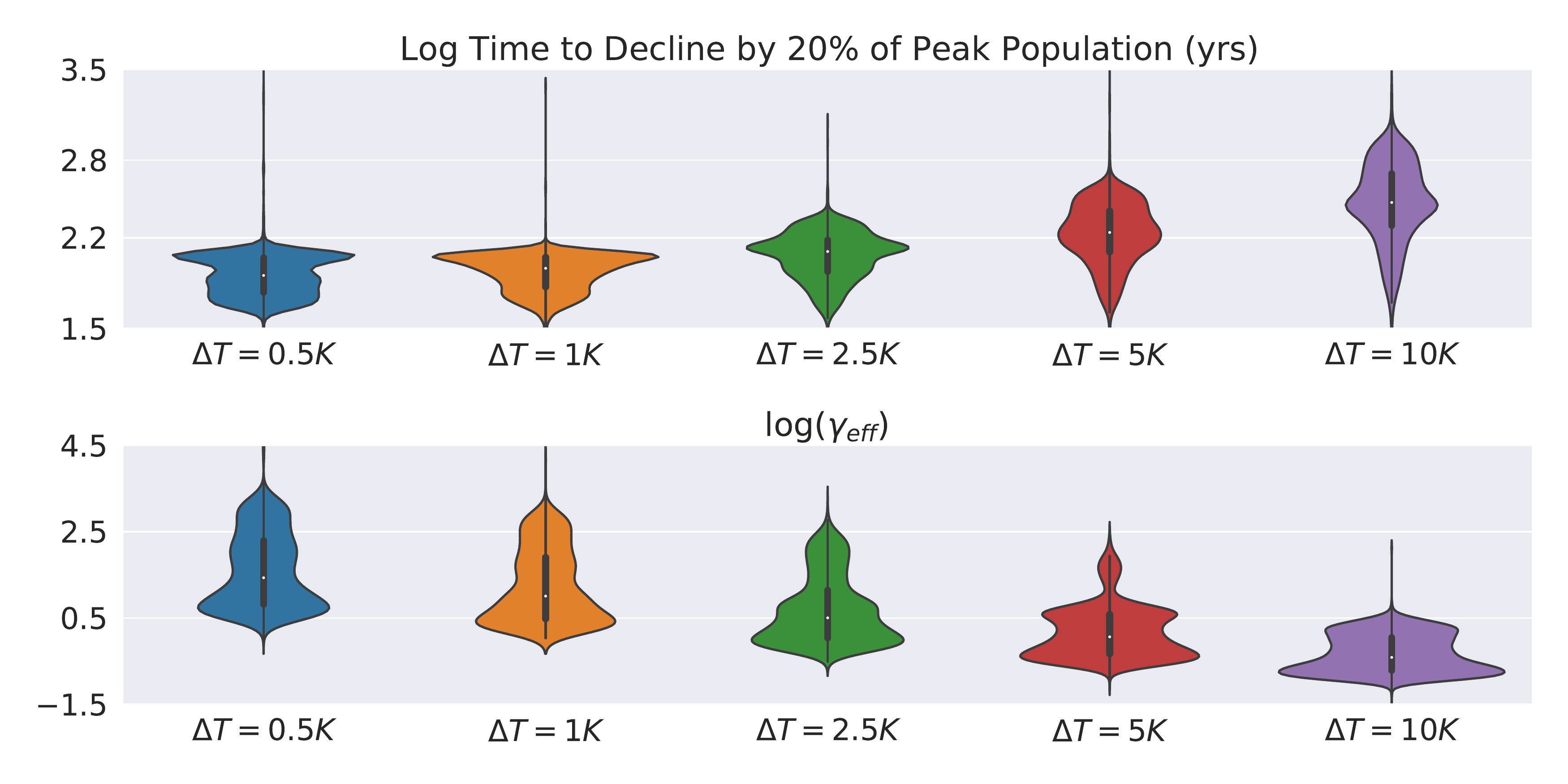}
    \caption{This figure shows a box plot of the marginal distributions shown in figure \ref{fig:scatterKDE}.  The white dot represents the median value.  The box's lower bound corresponds to the median of the lower half of the data-set, while the box's upper bound is the median of the upper half of the data-set.  It is of interest to note how increasing the population-temperature sensitivity parameter ($\Delta T$) results in a steadily decreasing $\gamma$ and a steadily increasing time for the population to decline.}
    \label{fig:violinPlots}
\end{figure*}

To see the effect of $\Delta T$ on the models, Figure \ref{fig:violinPlots} shows the marginal probability distributions for both population decline times and for $\gamma$.  These are represented as "violin plots" showing the range of values for both for runs with a given $\Delta T$.  These plots also show the  average and first moment of the distributions.

Note that the median values for the collapse times increase with $\Delta T$. This is to be expected as the sooner the planetary temperature rises beyond $T_0+\Delta T$ the sooner the population growth is truncated by climate effects.   The values of $\gamma$ extracted from the models reflect this, showing a decrease with increasing temperature tolerance.  Note that for $\Delta T < 5K$, the decline times are less than, or of order, a century.  Given that the timescale for a generation in our models is $\sim 25$ y, this means significant population loss across the lifetime of an individual.  We consider such a situation as likely to pose the greatest risk for civilizations.

\begin{figure*}
    \centering
    \includegraphics[scale=0.475]{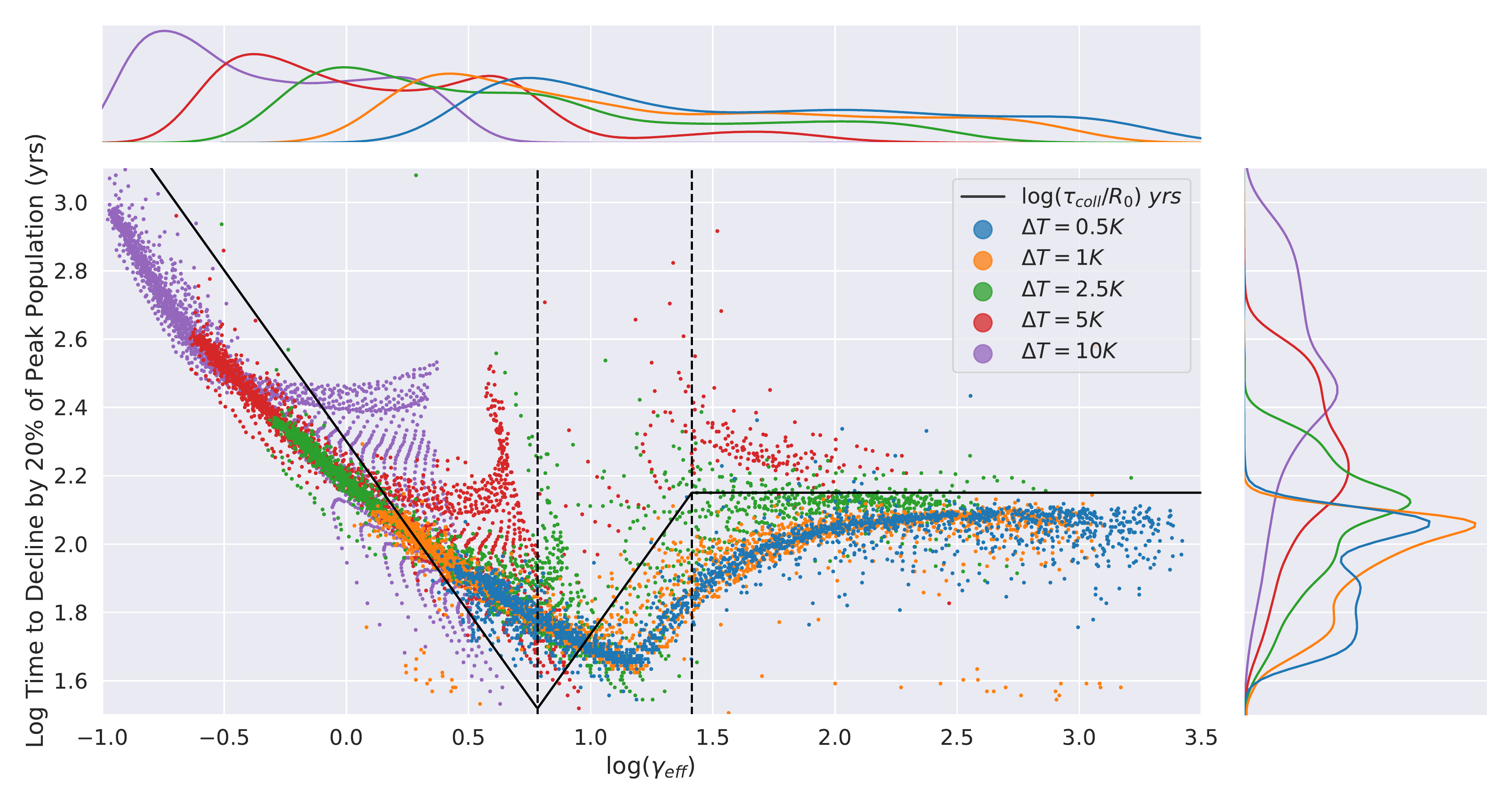}
    \caption{Shown above is a scatter plot of gamma versus the time for civilizations to decline by 20\% of the their peak population.  The solid black line shows our analytically derived collapse time, given by equation ($\ref{eq:tcoll}$).  The colors represent different values of our population-temperature tolerance parameter $\Delta T$.  The black dotted lines divide the graph into three regions.  The leftmost region corresponds to low $\gamma$ and high $\theta$.  This region results largely in overpopulation, which is the reason behind the long decline times.  The middle region corresponds to both high $\gamma$ and high $\theta$, where technology is able to accelerate birth rates, but eventually ends up contributing to an increased death rate and short decline time.  The rightmost region corresponds to a high $\gamma$ and a low $\theta$, where climate changes on a faster timescale then both technology and population growth.  Thus, in this region, civilizations experience climate change before they experience any growth benefits due to technology.  This means that these civilizations reach only a tiny fraction of their carrying capacity before they begin to decline.}
    \label{fig:scatterKDE}
\end{figure*}

We also show data for the models as a scatter plot of gamma versus decline time in Figure \ref{fig:scatterKDE}.  On the $x$ and $y$ axis we show corresponding marginal probability distributions for $\gamma$ and the decline time.  These are shown as a kernel density estimation graph such that the area under each curve is normalized to one. 

Shown as the solid black line is our analytically derived collapse time, given by equation (\ref{eq:tcoll}).  It can be seen that our analytic approximation reproduces the necessary features that arise in our numerical runs.  The deviations are explained in Figure \ref{fig:scatterGrid} in Appendix $A$.

There are three distinct regions for $\tau_{coll}$.  Starting from the right hand side of Figure \ref{fig:scatterKDE}, we are in the region of high $\gamma$ and low $\theta$.  We say that civilizations in this region experience a climate-dominated anthropocene.  The high $\gamma$ indicates that the timescale for the climate to change is much shorter than the timescale for population growth.  The low $\theta$ indicates that the timescale for climate to change is also much shorter than the timescale for technology to evolve.  Thus, in this region, civilizations experience climate change before they experience any significant growth benefits due to their technological capacities.  Thus these civilizations reach only a tiny fraction of their carrying capacity before they begin to decline.  Since they do not have much time to increase their overall growth rate above the "natural" value, their collapse rate ends up being approximately equivalent in magnitude to their natural growth rate.

As we move leftward in Figure \ref{fig:scatterKDE} to lower $\gamma$, we see a dip in the collapse times (this is the valley seen in the contour plots discussed in the last section).  This region, between the two black dotted lines, corresponds to high $\gamma$ and high $\theta$.  We say civilizations in this region experience a technology-dominated Anthropocene.  The high $\gamma$ indicates that the timescale for climate to change is much shorter than the timescale for population growth.  The high $\theta$ indicates that the timescale for climate to change is now longer than the timescale for technology to evolve.  Thus civilizations in this range experience a birth benefit due to their technological capabilities {\it before} they experience adverse effects stemming from climate change.  This is the region Earth currently inhabits.  The technological birth benefit allows the civilization to begin approaching their planet's carrying capacity.  However the high $\gamma$ value prohibits them from reaching this limit.  As we move towards the leftmost vertical dotted black line in the plot, civilizations peak closer to their carrying capacity.  Since the decline time for civilizations correlates to their growth rate, civilizations whose populations begin falling closer to their carrying capacity experience the fastest decline times. In essence, in this region the more technology props a civilization up, the harder they end up falling.

Continuing leftward in Figure \ref{fig:scatterKDE} to lower $\gamma$, we begin to leave the region of the Anthropocene and approach the limit of overpopulation.  In this region, $\theta$ is much greater than one, while $\gamma$ is now decreasing $< 1$.  Low $\gamma$ means that the timescale for population to grow is now much shorter than the timescale for the climate to change.  High $\theta$ again indicates that the timescale for climate to change is also much longer than the timescale for technology to evolve.  Thus, in this region, civilizations experience a birth benefit due to their growing technological capacities, allowing them to rise in population towards their carrying capacity.  As $\gamma$ drops below 1 civilizations are able to reach their carrying capacities ($\eta\to 1$).  In this region, the timescale for these civilizations to decline is then dictated by the timescale for climate to change, thus resulting in a longer collapse time $\tau_{coll}=1/\gamma=t_C/t_G$.  Moving further leftward yields ever lower values of $\gamma < 1$ and the timescale for climate to change becomes larger resulting in a steadily increasing collapse time. 

Finally, we note that in Figure \ref{fig:scatterKDE} the marginal distribution of the decline time for $\Delta T=1K$ seems to peak higher than $\Delta T=0.5K$.  This occurs because many of the $\Delta T=0.5K$ models are in the valley of our collapse time, and the marginal distribution is a projection onto the $y$-axis.  This can also be seen as the large shoulder protruding in the distribution of collapse times for $\Delta T=0.5K$.

The most important takeaway from these results is that an ever increasing share of the models experience climate dominated anthropocenes as $\Delta T$ decreases.  For $\Delta T < 5K$, most models experience rapid population declines. Even for $\Delta T = 10K$, the average decline time was $384$ years and $22.2\%$ of models had decline times less than 200 years.

\section{Discussion and Conclusions}

In this paper we have modeled the coupled evolution of a planet and a civilization through the era when energy harvesting by the civilization drives the planet into new and potentially adverse climate states. Our goal, continuing from the work of \cite{Frank2018}, is to determine if "anthropocenes" of the kind humanity is experiencing now might be a generic feature of planet-civilization evolution \citep{Haqq-Misra2009,Frank2017,MullanHaqq-Misra2019}. To this end we introduced and analyzed a set of coupled ODE's that track the trajectory of the civilization's population $(N)$, the generation of $CO_2$ via the civilization's energy harvesting activities and, finally, the mean planetary temperature $T$.  The principle innovation in this paper over \citep{Frank2018} is the use of an energy balance model to track the change in climate state as the atmospheric composition, $P=pCO_2$, changes due to the civilization's growth.  A different form for the population growth equation compared to \citep{Frank2018} was also used.

Using both direct simulation of the non-linear model and an analysis of a linearized, non-dimensional version of the equations we have found that most planets in the Complex Life Habitable Zone (CLHZ \cite{Schwietermanea2019}) undergo a {\it climate-dominated anthropocene}.  We define this to be a trajectory of the coupled planet-civilization system in which population growth is truncated by changes in the climate state driven by the civilization. Our analysis further shows such climate-dominated anthropocenes can occur in different ways depending on the planets initial atmospheric composition, orbital location and the technological capacities of the civilization.  On planets with high climate sensitivity ($dT/dP$), even modest technological innovation in terms of energy harvesting capacities can trigger detrimental climate change (Figure \ref{figs:alpha}). This situation is likely to occur at the inner edge of the habitable zone and would be akin to climate change occurring during the early era of industrialization on Earth (i.e. the Victorian period in England).  Planets on the outer edge of the host star's habitable zone require higher values of $CO_2$ concentrations to be temperate.  Such worlds have lower climate sensitivity and civilizations there can drive higher fractional increases in $pCO_2$ before temperatures rise significantly. The technological capacities of the civilization effect the trajectories when they allow for population growth rapid enough that it can compete with increasing death rates from an adversely changing climate.  Finally, the tolerance of the civilizations to rising temperatures $\Delta T$ represents another important input condition and, as could be expected, lower values of $\Delta T$ led to stronger climate dominated anthropocenes (Figure \ref{fig:violinPlots}).

The models presented in this paper represent an advance over our initial study \citep{Frank2018} because they assumed a specific form of energy harvesting  (combustion) and included an explicit physical models for its climate impact via a combustion dependent (i.e. $CO_2$) Energy Balance Model.  In this way we added a higher level of physio-chemical realism to the planet-civilization model system. With the new model we were able to vary parameters such as orbital distance and atmospheric composition for a specific class of worlds (i.e. Earth-like planets orbiting Sun-like stars).

One significant question to arise from our studies is the applicability of the Complex Life Habitable Zone \citep{Schwietermanea2019,Ramirez2020}.  For evolutionary reasons it is generally believed that a technological civilization would only arise from complex multi-cellular "animal" life \citep{Carter2008,Watson2008}.  \cite{Schwietermanea2019} emphasized that the oxidation of organic matter via free $O_2$ is the best means of producing significant free energy via respiration.  Given that $O_2$ is the only high-potential oxidant sufficiently stable to accumulate within planetary atmospheres \citep{Catlingea2005} it comprises a necessary condition for intelligence species.  Based on Earth's evolutionary history it is clear that complex aerobic life can also be strongly impacted by $CO_2$ with limits for animal life appearing at atmospheric conditions of $pCO_2 > 5\times10^4$ ppm (humans $pCO_2$ lethality limit may be 10 times lower).  Thus, if we take Earth's history as a guide, atmospheres with high oxygen and low $CO_2$ may be necessary for the emergence of intelligent civilization-building species.  If this is the case then climate-dominated anthropocenes would appear to be a generic feature of coupled planet-civilization evolution.  The difficulty however is knowing how universal the constraints on $pCO_2$ are for life elsewhere.  Is it possible for a planet with $10^5$ ppm of $CO_2$ to evolve complex life that goes on to create a technological civilization?  In addition, what constraints exist for energy harvesting based on combustion on high $pCO_2$ worlds?  If civilizations could occur on such planets then our results indicate that these worlds would not undergo a climate-dominated anthropocene.  Instead, in our models, their populations climb until they reach the planet's carrying capacity.  This situation, however, presents its own difficulties and could represent a different form of negative feedback on the planet-civilization system that is not modeled in our equations.  It is also possible that without rapid detrimental feedbacks from adverse climate change a civilization would reduce its population growth on its own before carrying capacity is reached. Finally we note that \citet{Howell2019} criticized the use of the  CLHZ as being too narrow a a view of evolutionary possibilities.  We view certainly has merit we however that the criticism was based mainly an arguments surrounding the biochemistry of $CO$ and not $CO_2$.

Future modeling efforts could explore different representations of the population growth and its coupling to climate state and energy harvesting modality.  In addition, other energy harvesting modalities such as wind or solar could also be explored.  Future work should also investigate the impact of the stellar spectral type, as this will impact the climate of the planet and may also impact the ability of a civilization to utilize stellar energy (as M-dwarfs have a different SED than G-dwarfs, etc.). Finally models that include changes in the civilizations behavior, such as simply switching from one harvesting mode to another could also be included in the modeling.  Each of these steps would represent a further articulation of an {\it astrobiology of the anthropocene} by exploring issues associated with the biospheric dimensions of creating a long-term sustainable civilization.

\section*{Appendix A: Intrinsic Timescales and Dimensionless Quantities}
\label{sec:appA}
In analyzing the model we see that it contains three intrinsic timescales.
\begin{align*}
t_G&\equiv\frac{1}{R_0}=\textit{Timescale for Population Growth}\\
t_T&\equiv\frac{\Delta P}{CN_{max}}=\textit{Timescale for Technological Advancements}\\
t_C&\equiv\frac{\Delta T}{CN_{max}D}=\textit{Timescale for Climate Change}
\end{align*}
Where \[D\equiv \frac{\Delta T_F}{P_0}=\frac{dT}{dP}\bigg|_{T=T_0}=\textit{Initial Climate Sensitivity}\]
as defined in equation $(\ref{eq:climateSensitivity})$.  Also, $R_0$ is the initial "natural" growth rate, $\Delta P$ is the technological birth benefit, $\Delta T$ is the civilizations temperature tolerance, $N_{max}$ is the carrying capacity and $C$ is the annual, per-capita carbon footprint.  We can use these timescales to define our dimensionless quantities shown in Table \ref{tab:dimQs}.
\begin{equation*}
    \gamma\equiv\frac{t_G}{t_C}=\frac{DCN_{max}}{R_0\Delta T}\quad
    \theta\equiv\frac{t_C}{t_T}=\frac{\Delta T}{D\Delta P}\quad
    \beta\equiv\frac{t_G}{t_T}=\frac{CN_{max}}{R_0\Delta P}
\end{equation*}

\begin{figure*}[t]
    \centering
    \includegraphics[width=\textwidth]{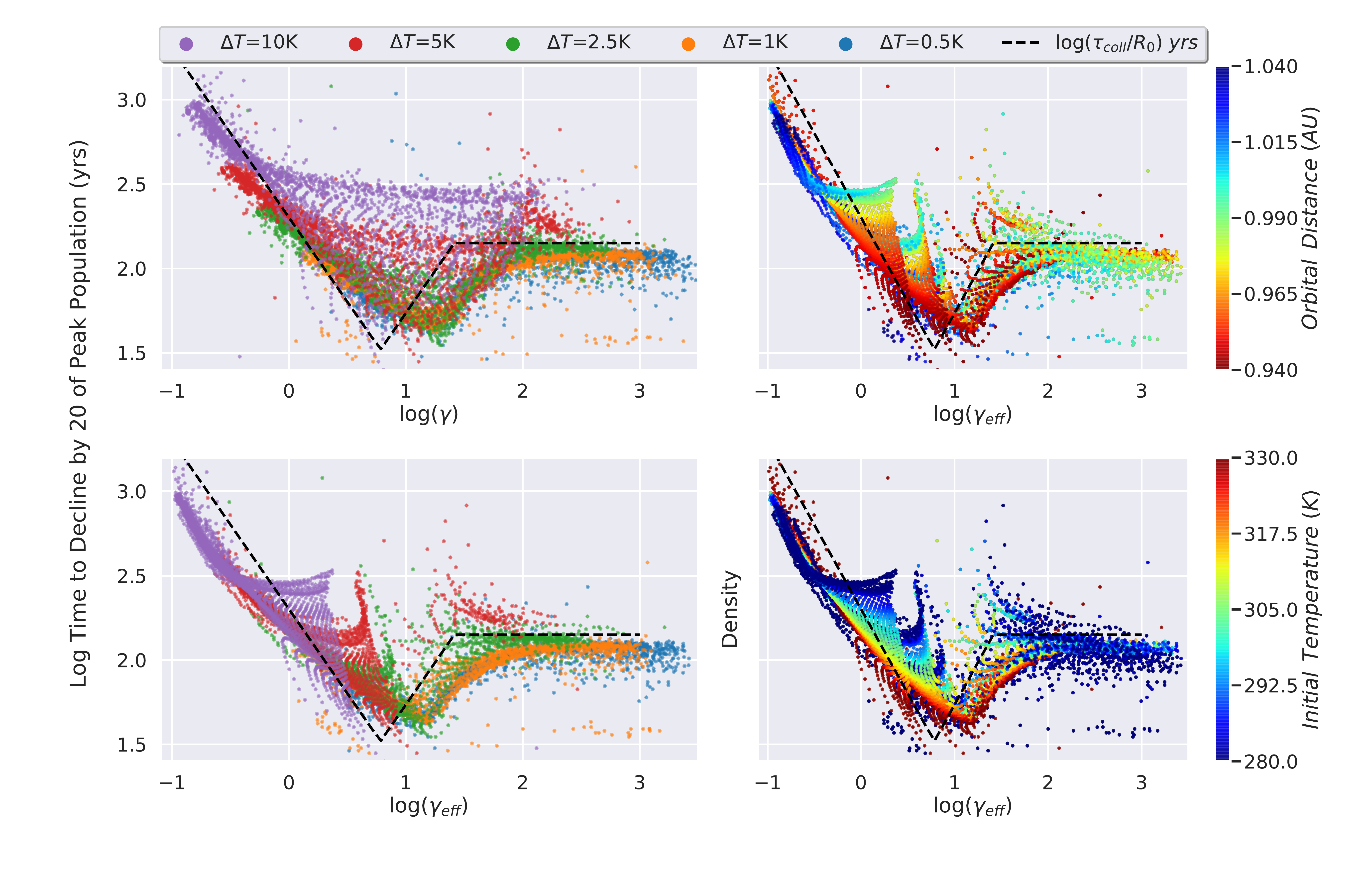}
    \caption{The top left plot shows our numerically calculated values of $\gamma$, as defined in table \ref{tab:dimQs} and derived in appendix A, plotted versus the numerically calculated times for our models populations to decline by $20\%$ from their peak values.  The black dotted line shows our analytical expression for collapse times, derived in appendix B.  For contrast, the bottom left plot shows the same things yet instead for $\gamma_{\text{eff}}$, as given by equation \ref{eq:gammaEff}.  The biggest difference as compared to the plot above is a net decrease in $\gamma$, which is as expected from equation \ref{eq:gammaCompare}.  This also has the effect of greatly reducing the deviations from our analytical predicitoins, that is, adheres much more to our check-mark shaped prediction.  Although, not all deviations have been solved by using $\gamma_{\text{eff}}$.  The right column shows plots of $\gamma_{eff}$ colored by orbital distance (top), and initial global temperature (bottom).  It can be seen that the models that deviate greatest from our analytical prediction or those that have large orbital radii and low initial global temperatures.  As shown in figure \ref{fig:paramSweepdT5}, the contour lines for $pCO_2$ travel diagonally across the parameter space of $a/T_0$.  Thus, as a result, civilizations with high orbital distance and low initial global temperature could have the same value of initial $pCO_2$ as the civilizations with low orbital distance and high initial global temperature.  Since $\gamma$ is principally dependent on the value of the initial $pCO_2$, this means that these two classes of civilizations will have the same value of $\gamma$ and hence the same value of $\tauColl$.  Although, in reality, the civilizations with the higher orbital radii end up taking longer to fall, hence have longer decline times.
}
    \label{fig:scatterGrid}
\end{figure*}

\noindent
Note that $\beta=\theta\gamma$.  Since we have defined $\gamma$ to be approximately constant, based on the initial climate sensitivity, we can define an 'effective' $\gamma$ to be based on the value of the climate sensitivity when the population reaches its peak value, which occurs approximately when global temperatures have deviated from their initial values by $\Delta T$.
\[\gamma_{\text{eff}}\equiv\left(\frac{CN_{max}}{R_0\Delta T}\right)\frac{dT}{dP}\bigg|_{T\approx T_0+\Delta T}\tag{A1}\label{eq:gammaEff}\]
As defined by equation \ref{eq:climateSensitivity}, this means that
\begin{align*}\gamma_{\text{eff}}&=\left(\frac{CN_{max}}{R_0\Delta T}\right)\frac{\Delta T_F}{P_0}e^{-\Delta T/\Delta T_F}\\&\equiv \left(\frac{DCN_{max}}{R_0\Delta T}\right)e^{-\alpha}=\gamma e^{-\alpha}\tag{A2}\label{eq:gammaCompare}\end{align*}
Where we have defined $\alpha\equiv \Delta T_F/\Delta T$.  Thus, as temperatures increase, the civilizations effective $\gamma$ will drop from its fiducial value.  See Figure \ref{fig:scatterGrid} to visualize this effect.  Furthermore, the value of the carrying capacity that makes $\gamma=1$ is indicative of the number of people required to force the climate out of equilibrium in a single growth timescale $(t_G)$.  This quantity is called the civilizations "anthropogenic population"
\[N_A\equiv\frac{\gamma}{N_{max}}=\frac{R_0\Delta T}{DC}=\textit{Anthropogenic Population}\]

\section*{Appendix B: Derivation of Collapse Time for High Climate Forcing}
\label{sec:appB}
For high climate forcing $(\gamma\gg1)$ we find that the collapse time is given by
\[\tauColl=\frac{1}{\max(\sqrt{2},\theta)}=\begin{cases}
1/\sqrt{2}&\theta\ll1\\
1/\theta & \theta\gg1
\end{cases}\tag{B1}\]
We can derive this from our linearized model as follows.  If $\gamma\gg1$, then $t_C<t_G$, so the climate changes on a much faster rate than the population does.  This is the region of the climate-driven Anthropocene, which means that global temperatures will begin to increase death rates before civilizations reach their carrying capacity.  A consequence of this is that civilizations in this region never reach their maximum growth rate, $\eta\ll1$, which implies that $1-\eta>\eta(1+\theta\epsilon)$, so their population growth equation will have the form.
\[\dot{\eta}=\eta(1+\theta\epsilon)-\eta\epsilon^2=\eta(1+\theta\epsilon-\epsilon^2)\equiv d\eta/d\tau\]
We can set $\dot{\eta}=0$ and solve for $\epsilon$ to determine what the environmental state is at the point when population begins to collapse (ie. the moment when $\dot{\eta}$ drops below zero).  We call this value of $\epsilon$ its critical value, which we find to be
\[\epsilon_c=\frac{\theta}{2}+\sqrt{1+\left(\frac{\theta}{2}\right)^2}\tag{B2}\label{eq:critEpsilon}\]
We can also find the time scale for the population to decline after it reaches this peak.  Since these civilizations do not reach their carrying capacity, their growth rate as they enter the Anthropocene dictates what their collapse rate will be directly after.  Thus, we use the free fall time for civilizations as their collapse rate. 
\[\tauColl=\sqrt{-\frac{\eta}{\ddot{\eta}}}\bigg |_{\dot{\eta}=0}\quad (\gamma\gg 1)\]
We can use our population growth rate equation to find this second derivative
\[\ddot{\eta}=\eta(\theta\dot{\epsilon}-2\epsilon\dot{\epsilon})\]
Where we let $\dot{\epsilon}\equiv d\epsilon/d\tau$.  When population begins to decline, global temperatures continue to rise exponentially.  Thus, at this point, the rate of increasing temperatures is approximately equal to its critical value $\dot{\epsilon_c}\approx \epsilon_c$.  Thus, it follows that at this point, directly after $\dot{\eta}=0$...
\[-\ddot{\eta}\approx \eta(2\epsilon_c^2-\theta\epsilon_c)\]
Thus, it follows that the collapse timescale for $\gamma\gg1$ is given by
\[\tauColl=\left(2\epsilon_c^2-\theta\epsilon_c\right)^{-1/2}\quad (\gamma\gg1)\]
This can further be broken down into two cases, dependent on $\theta$.

\begin{enumerate}[(1)]
    \item \underline{Low Technological Growth Acceleration} 
    
    If $\theta\ll 1$, then $\epsilon_c\rightarrow 1$ and
    \[\tauColl=1/\sqrt{2}\]

    In this case, the civilizations technological abilities are not advanced enough to increase their growth rates.  Also, the fastest timescale is that for climate change.  Since their rate of technological advancements is negligible and their rate of climate change exceeds that for population growth, these civilizations will have negligible population growth before entering their Anthropocene.  As a result, their growth rate as they enter the Anthropocene will be approximately equal to its initial "natural" value.  Thus, all civilizations in this area will also collapse with a constant "natural" collapse rate.

    \item \underline{High Technological Growth Acceleration} 
    
    If $\theta\gg 1$, then $\epsilon_c\rightarrow \theta$ and
    \[\tauColl=1/\theta\]
    
    In this case the civilizations technological abilities are able to accelerate their growth rates.  Thus, these civilizations enter their Anthropocene with an accelerated growth rate due to their technological abilities.  This accelerated growth leads to an accelerated decline, which means that these civilizations collapse at a faster than natural rate.

\end{enumerate}

\acknowledgments
This work was supported by NASA grant \#80NSSC20K0622.

\bibliographystyle{mnras}

\begin{thebibliography}{99}

\bibitem[Carroll-Nellenback et al.(2019)]{Carroll-Nellenbackea2019} Carroll-Nellenback, J., Frank, A., Wright, J., Scharf, C.\ 2019.\ The Fermi Paradox and the Aurora Effect: Exo-civilization Settlement, Expansion, and Steady States.\ The Astronomical Journal 158. doi:10.3847/1538-3881/ab31a3

\bibitem[Carter (2008)]{Carter2008}Carter, B. (2008). Five- or six-step scenario for evolution? International Journal of Astrobiology, 7(2), 177-182. doi:10.1017/S1473550408004023

\bibitem[Catling et al.(2005)]{Catlingea2005} Catling, D.~C., Glein, C.~R., Zahnle, K.~J., McKay, C.~P.\ 2005.\ Why O$_{2}$ Is Required by Complex Life on Habitable Planets and the Concept of Planetary ``Oxygenation Time''.\ Astrobiology 5, 415–438. doi:10.1089/ast.2005.5.415

\bibitem[Cohen (1995)]{Cohen1995}Cohen JE. Population growth and earth's human carrying capacity. Science. 1995 Jul 21;269(5222):341-6. doi: 10.1126/science.7618100.

\bibitem[Crutzen (2002)]{Crutzen2002} Crutzen, P.~J.\ 2002.\ The ``anthropocene''.\ Journal de Physique IV 12, 1–5. doi:10.1051/jp4:20020447

\bibitem[Elhacham et al. (2020)]{Elahacham2020}Elhacham, E., Ben-Uri, L., Grozovski, J. et al.\ 2020.\ Global human-made mass exceeds all living biomass \ Nature 588, 442–444, https://doi.org/10.1038/s41586-020-3010-5

\bibitem[Fairen et al. (2012)]{Fairen2012}A. G. Fairen and J. D. Haqq-Misra and C. P. McKay\ 2012\ Reduced albedo on early Mars does not solve the climate paradox under a faint young Sun \ EDP Sciences 540, A13, https://doi.org/10.1051/0004-6361/201118527

\bibitem[Frank Alberti \& Kleidon (2017)]{Frank2017} Frank, A., Alberti, M., \& Keliedon, A.,. 2017, "Earth as a Hybrid Planet, The Anthropocene in an Evolutionary Astrobiological Context, Anthropocene,  (in press).

\bibitem[Frank et al.(2018)]{Frank2018} Frank, A., Carroll-Nellenback, J., Alberti, M., Kleidon, A.\ 2018.\ The Anthropocene Generalized: Evolution of Exo-Civilizations and Their Planetary Feedback.\ Astrobiology 18, 503–518. doi:10.1089/ast.2017.1671

\bibitem[Gaidos \& Knoll (2012)]{GaidosKnoll2012} Gaidos, E., Knoll, A.H., 2012. Our evolving planet: from the dark ages to an evolutionary renaissance. In: Impey, C., Lunine, J., Funes, J. (Eds.), Frontiers in Astrobiology. Cambridge University Press, Cambridge.

\bibitem[Haqq-Misra and Baum(2009)]{Haqq-Misra2009} Haqq-Misra, J.~D., Baum, S.~D.\ 2009.\ The Sustainability Solution To The Fermi Paradox.\ Journal of the British Interplanetary Society 62, 47-51. 


\bibitem[Huang \& Bani Shahabadi (2014)]{HuangBani2014}Huang, Y., Bani Shahabadi, M. \ 2014. \  Why logarithmic? A note on the dependence of radiative forcing on gas concentration \ J. Geophys. Res. Atmos., 119, 13,683– 13,689, doi:10.1002/2014JD022466

\bibitem[IPCC (2014)]{IPCC2014}Climate Change 2014: Synthesis Report. Contribution of Working Groups I, II and III to the Fifth Assessment Report of the Intergovernmental Panel on Climate Change [Core Writing Team, R.K. Pachauri and L.A. Meyer (eds.)]. IPCC, Geneva, Switzerland, page 43

\bibitem[Mullan and Haqq-Misra(2019)]{MullanHaqq-Misra2019} Mullan, B., Haqq-Misra, J.\ 2019.\ Population growth, energy use, and the implications for the search for extraterrestrial intelligence.\ Futures 106, 4–17. doi:10.1016/j.futures.2018.06.009


\bibitem[Hooke \etal (2013)]{Hooke2013} Hooke,  R., Martín-Duque, J.,  Pedraza, J. (2012) Land transformation by humans: A review GSA Today, 22(12), 5- 10

\bibitem[Howard (2013)]{Howard2013} Howard, A.W. (2013) “Observed Properties of Extrasolar Planets”. Science 340, 572-576.

\bibitem[Howell (2019)]{Howell2019} Howell, J.~E., 2019 "Comment on "A Limited Habitable Zone for Complex Life", Res. Notes AAS 3 85

\bibitem[Jones(1976)]{1976Icar...28..421J} Jones, E.~M.\ 1976.\ Colonization of the Galaxy.\ Icarus 28, 421. 

\bibitem[Kot (2001)]{Kot2001}Kot, M., 2011, Elements of Mathematical Ecology, Cambridge.

\bibitem[Kleidon (2010)]{Kleidon2010 } Kleidon, A. (2010) Life, hierarchy, and the thermodynamic machinery of planet Earth. Physics of Life Reviews, 7, 424-460.

\bibitem[Kleidon (2012)]{Kleidon2012} Kleidon, A., 2012. How does the Earth system generate and maintain thermodynamic disequilibrium and what does it imply for the future of the planet?. Phil. Trans. R. Soc. A 370: 1012-1040.

\bibitem[Kennett \& Beach (2013)]{KennettBeach2013} Kennett D.,\& Beach, T. P. (2013) Archeological and environmental lessons for the Anthropocene from the Classic Maya collapse Anthropocene 4 (2013) 88–100

\bibitem[Korpela  (2015)]{Korpella2015} Korpela, E. J., Sallmen, S. M., \& Leystra Greene, D. 2015, ApJ

\bibitem[Landis(1998)]{1998JBIS...51..163L} Landis, G.~A.\ 1998.\ The Fermi paradox: an approach based on percolation theory..\ Journal of the British Interplanetary Society 51, 163-166. 

\bibitem[Lenton \etal (2008)]{Lenton2008} Lenton, T, Held, H., Kriegler, E., Hall, H., Lucht, W., Rahmstorf, S., Schellnhuber, H., 2008, PNAS, 105, 1786-1793, ``Tipping elements in the Earth’s climate system"

\bibitem[Lin (2014)] {LinetalApJL2014}Lin H. W., Gonzalez Abad G., Loeb A., 2014, ApJL, 792, L7 

\bibitem[Lingam and Loeb(2017)]{2017MNRAS.470L..82L} Lingam, M., Loeb, A.\ 2017.\ Natural and artificial spectral edges in exoplanets.\ Monthly Notices of the Royal Astronomical Society 470, L82-L86. 

\bibitem[Kuehn (2011)]{Kuehn2011} Kuehn,C., Physica D, 2011, 240  1020–1035 ``A mathematical framework for critical transitions: Bifurcations, fast–slow systems and stochastic dynamics"

\bibitem[Lenton (2011)]{Lenton 2011} Lenton, T., 2011, Nature Climate Change 1, 201–209, ``Early warning of climate tipping points"

\bibitem[Lenton \etal (2015)]{Lenton2015} Lenton et al. 2015 Planetary boundaries: Guiding human development on a changing planet. Science, 347, 736

\bibitem[Lenton, Pichler \& Weiz (2016)]{Lenton2016} Lenton, T.M., Pichler, P., \& Weisz H.,Earth Syst. Dynam., 7, 353-370, 2016

\bibitem[Lingam and Loeb(2019)]{LingamLoeb2019} Lingam, M., Loeb, A.\ 2019.\ Relative Likelihood of Success in the Search for Primitive versus Intelligent Extraterrestrial Life.\ Astrobiology 19, 28–39. doi:10.1089/ast.2018.1936


\bibitem[Meadows {\it et al.} (1972)]{Meadows72} D. H. Meadows and D. L Meadows and J. Randers and W. W. Behrens III, The limits to growth, 1972, Universe Books

\bibitem[Mattews \etal (2011)]{Mattews2011} Matthews, B., A. Narwani, S. Hausch, E. Nonaka, H. Peter, M. Yamamichi, K. Sullam, K. Bird, M. Thomas, T. Hanley, and C. Turner. (2011). Toward an integration of evolutionary biology and ecosystem science. Ecology Letters 14(7): 690-701.
1
\bibitem[Miller \etal (2011)]{Miller2011} Miller1 L., Gans, F., Kleidon, A., 2011, Earth Syst. Dynam., 2, 1–12, ``Estimating maximum global land surface wind power extractability
and associated climatic consequences"

\bibitem[North \& Kim (2017)]{NorthKim2017} North, G \& Kim, K., 2017, Energy Balance Climate Models, Wiley, New York.

\bibitem[Newman and Sagan(1981)]{1981Icar...46..293N} Newman, W.~I., Sagan, C.\ 1981.\ Galactic civilizations - Population dynamics and interstellar diffusion.\ Icarus 46, 293-327. 

\bibitem[Perretto \& Valente (2015)]{PerrettoValente2015} Perretto, P., \& Valente S., 2015, J. Econ Growth, 20,305-331

\bibitem[Pimentel (1961)]{Pimentel 1961} Pimentel, D. (1961) Animal population regulation by the genetic feedback mechanism. Am. Nat. 95, 65–79

\bibitem[Ramirez(2020)]{Ramirez2020} Ramirez, R.~M.\ 2020.\ A Complex Life Habitable Zone Based On Lipid Solubility Theory.\ Scientific Reports 10. doi:10.1038/s41598-020-64436-z

\bibitem[Reuvey (2012)]{Reuvey2012} Reuveny, R., 2012, Annu. Rev. Resour. Econ. 2012. 4:303–264

\bibitem[Rockstrom \etal (2009)]{Rockstrom2009}Rockstrom, J. et al. (2009) Planetary boundaries: exploring the safe operating space for humanity. Ecol. Soc. 14, 32

\bibitem[Rull (2016)]{Rull2016}Rull, V., 2016, Quaternary Science Reviews 150 (2016) 31e41

\bibitem[Seager (2013)]{Seager2013}Seager, S., 2013. ExoPlanet habitability. Science 340, 577–581.

\bibitem[Schellnhuber (1998)] {Schellnhuber1998} Schellnhuber, H.J., Wenzel, V, Earth System Analysis, Integrating Science for Sustainability, 1998, Editors Schellnhuber, H.J., Wenzel, V., Springer 

\bibitem[Solomon \etal (2007)]{Solomon2007} Solomon, S., Qin, D., Manning, M., Chen, Z., Marquis, M., Averyt, K.B., Tignor, M., Miller, H.L. (Eds.), 2007. Climate Change 2007: The Physical Science Basis. Cambridge University Press, Cambridge, United Kingdom and New York, NY,
USA.

\bibitem[Sufan \etal (2014)]{Sufanea2014}Safuan, H.,Sidhu, H., Jovanoski, Z., Towers, I., 2014, ANZIAM J., 54, pp.C768–C787

\bibitem[Schwieterman et al.(2019)]{Schwietermanea2019} Schwieterman, E.~W., Reinhard, C.~T., Olson, S.~L., Harman, C.~E., Lyons, T.~W.\ 2019.\ A Limited Habitable Zone for Complex Life.\ The Astrophysical Journal 878. doi:10.3847/1538-4357/ab1d52

\bibitem[Solomon \etal (2007)]{Solomon2007} Solomon, S., D. Qin, M. Manning, Z. Chen, M. Marquis, K.B. Averyt, M. Tignor and H.L. Miller (eds.) Contribution of Working Group I to the Fourth Assessment Report of the Intergovernmental Panel on Climate Change, 2007, Cambridge University Press, Cambridge, United Kingdom and New York, NY, USA.

\bibitem[Vitousek \etal (1986)]{VitousekEtal1986} Vitousek, P. M. Ehrlich, P. R., Ehrlich, A. H., Matson, P. A., 1986. Human appropriation of the products of photosynthesis. Bioscience 36: 368-373.


\bibitem[Watson \& Lovelock (1983)]{WatsonLovelock1983} A. J. Watson and J. E. Lovelock, 1983, Biological homeostasis of the global environment: the parable of Daisyworld, Tellus, 35B,284-289 

\bibitem[Way \etal (2017)]{Way2017} Way, M.~J., and 10 colleagues 2017.\ Resolving Orbital and Climate Keys of Earth and Extraterrestrial Environments with Dynamics 1.0: A General Circulation Model for Simulating the Climates of Rocky Planets.\ ArXiv e-prints arXiv:1701.02360.

\bibitem[Williams \& Kasting (1997)]{Kastings1997} Williams,  Darren  M.  and  James  F.  Kasting  (1997)   “Habitable  Planets  with  High Obliquities”. In:Icarus129.1, pp. 254 –267. ISSN: 0019-1035.

\bibitem[Williamson \etal (2016)] {Williamson2106 } Williamson, M., Bathiany, S.,Lenton, T, 2016, Earth Syst. Dynam., 7, 313–326, 2016, ``Early warning signals of tipping points in periodically forced systems"

\bibitem[Wright et al.(2016)]{WrightetalApJ2016} Wright, J.~T., Cartier, K.~M.~S., Zhao, M., Jontof-Hutter, D., Ford, E.~B.\ 2016.\ The   Search for Extraterrestrial Civilizations with Large Energy Supplies. IV. The Signatures and Information Content of Transiting Megastructures.\ The Astrophysical Journal 816, 17.

\bibitem[Watson (2008)]{Watson2008}Carter, B. (2008) Watson, A, 2008, Implications of an Anthropic Model of Evolution for
Emergence of Complex Life and Intelligence, ASTROBIOLOGY
Volume 8, Number 1, 2008

\bibitem[Wittmann \& Portner (2013) ]{WittmannPortner2013} Wittmann, A. C., \& Pörtner, H.-O. 2013, Sensitivities of extant animal taxa to ocean acidification, NatCC, 3, 995


\bibitem[Zhang (2015)]{Zhang2015 }Zhang, X., Christian Kuehn, C., Sarah Hallerberg, S. 2015, PHYSICAL REVIEW E 92, 052905, ``Predictability of critical transitions"
\end{thebibliography}

\end{document}